\documentclass[11pt, oneside, reqno]{amsart}

 \usepackage{curves} 
\usepackage{graphicx} 
\usepackage{amsmath, amsfonts, amssymb, amscd,graphics, graphpap, tikz, color, xcolor, cancel,enumerate,enumitem, hyperref, mathabx, mathrsfs, mathtools}
\usepackage[scr=rsfs,cal=boondox]{mathalfa}

\usepackage{geometry}
\usetikzlibrary{arrows.meta,arrows}
\usetikzlibrary{decorations.pathreplacing,decorations.markings}
\tikzset{arrow data/.style 2 args={%
      decoration={%
         markings,
         mark=at position #1 with \arrow{#2}},
         postaction=decorate}
      }%
      
\usepackage[mathscr]{euscript}
  \usepackage[normalem]{ulem}
  \usepackage{soul}

\numberwithin{equation}{section}

\theoremstyle{plain}
\newtheorem{theo}{Theorem}[section]
\newtheorem{lem}[theo]{Lemma}
\newtheorem{prop}[theo]{Proposition}

\newtheorem{cor}[theo]{Corollary}

\theoremstyle{definition}
\newtheorem{rem}[theo]{Remark}

\newtheorem{example}[theo]{Example}
\newtheorem{definition}[theo]{Definition}

\newenvironment{pf}{\noindent{\it Proof. }}{$\hfill\square$\par\medskip}

\theoremstyle{plain}

\theoremstyle{definition}

\newcommand{\beq}{\begin{equation}}
\newcommand{\eeq}{\end{equation}}
\newcommand{\beqn}{\begin{equation*}}
\newcommand{\eeqn}{\end{equation*}}
\renewcommand{\a}{\alpha}
\renewcommand{\b}{\beta}

\renewcommand{\d}{\delta}

\newcommand{\ve}{\varepsilon}
\newcommand{\f}{\varphi}
\newcommand{\g}{\gamma}

\renewcommand{\k}{\kappa}
\newcommand{\lm}{\lambda}

\renewcommand{\r}{\rho}

\newcommand{\x}{\xi}

\renewcommand{\O}{\Omega}

\newcommand{\bC}{\mathbb{C}}
\newcommand{\bD}{\mathbb{D}}

\newcommand{\bR}{\mathbb{R}}

\newcommand{\bT}{\mathbb{T}}

\renewcommand{\gg}{\mathfrak{g}}

\newcommand{\gr}{\mathfrak{r}}

\newcommand{\cB}{\mathscr{B}}
\newcommand{\cC}{\mathcal{C}}
\newcommand{\cD}{\mathscr{D}}

\newcommand{\cI}{\mathscr{I}}
\newcommand{\cJ}{\mathscr{J}}

\newcommand{\cM}{\mathscr{M}}

\newcommand{\cS}{\mathscr{S}}
\newcommand{\cT}{\mathscr{T}}
\newcommand{\cU}{\mathscr{U}}
\newcommand{\cV}{\mathscr{V}}

\newcommand{\sC}{\mathsf{C}{}}
\newcommand{\sT}{\mathscr{T}{}}

\newcommand{\p}{\partial}

\renewcommand{\square}{\kern1pt\vbox
{\hrule height 0.6pt\hbox{\vrule width 0.6pt\hskip 3pt
\vbox{\vskip 6pt}\hskip 3pt\vrule width 0.6pt}\hrule height0.6pt}\kern1pt}

\DeclareMathOperator{\Span}{span}

\renewcommand\={:=}

\newcommand{\wt}{\widetilde}

\newcommand{\bt}{\begin{theo}}
\newcommand{\et}{\end{theo}}
\newcommand{\bp}{\begin{prop}}
\newcommand{\ep}{\end{prop}}
\newcommand{\bc}{\begin{cor}\ \ }
\newcommand{\ec}{\end{cor}}
\newcommand{\bl}{\begin{lem}\ \ }
\newcommand{\el}{\end{lem}}
\newcommand{\bd}{\begin{definition}}
\newcommand{\ed}{\end{definition}}
\newcommand{\be}{\begin{equation}}
\newcommand{\ee}{\end{equation}}

\def\<#1,#2>{\langle\,#1,\,#2\,\rangle}
\newcommand{\arr}{\begin{array}{rlll}}
\newcommand{\ea}{\end{array}}
\newcommand{\bea}{\begin{eqnarray}}
\newcommand{\eea}{\end{eqnarray}}
\newcommand{\bean}{\begin{eqnarray*}}
\newcommand{\eean}{\end{eqnarray*}}

\font\smallsmc = cmcsc9
\font\smalltt = cmtt8
\font\smallit = cmti8

\def\sideremark#1{\ifvmode\leavevmode\fi\vadjust{
\vbox to0pt{\hbox to 0pt{\hskip\hsize\hskip1em
\vbox{\hsize2cm\tiny\raggedright\pretolerance10000
\noindent #1\hfill}\hss}\vbox to8pt{\vfil}\vss}}}

\newcommand{\I}{\operatorname{i}}

\newcommand\bex{\begin{example}}
\newcommand\eex{\end{example}}
\newcommand\br{\begin{rem}}
\newcommand\er{\end{rem}}

\makeatletter
\@namedef{subjclassname@1991}{2020 Mathematics Subject Classification}
\makeatother

\title[Black holes and  black regions, horizons and barriers]{Black holes and  black regions,
horizons  and barriers\\ in  Lorentzian manifolds}
\author{Cristina Giannotti}
\author{Andrea Spiro}

\begin{document}
\setcounter{section}0

\begin{abstract}  We prove that if $\cS$ is a time-oriented null hypersurface of  a Lorentzian $n$-manifold $(M, g)$, the causal world-lines,   which intersect transversally $\cS$ and  are time-oriented in a compatible 
way,   cross the hypersurface  all in the same direction, the other being forbidden. Even if  it is  known  that a smooth event horizon (in the sense of  Penrose, Hawking and Ellis)  is a null hypersurface and 
has the above  semi-permeability property,  at the best of our knowledge, in the literature it was not
 stated  so far  that the latter  is a mere consequence of the former.   Our result  leads  to  the  concepts  of  barriers (= null hypersurfaces separating the space-time into  disjoint regions)  and black regions (= time-oriented regions bounded by barriers). These objects    naturally  include  (smooth) event horizons and   (smoothly bounded) black holes.  
 Since  barriers are defined  by   two  simple  properties -- the merely local property of ``nullity''  combined with the global property of ``separating the space-time'' --   we  expect   they  may be used to  simplify   computations for   locating  static and/or dynamic  horizons in numerical computations. 
 \end{abstract}

\maketitle

\section{Introduction}

The  main result  of this paper consists of a   rigorous proof of the following property of   the  time-oriented  null hypersurfaces:  {\it any causal  world-line of  a Lorentzian manifold $(M, g)$,  which   intersects transversally  a time-oriented null hypersurface $\cS$   at a point $x_o$ and  is time-oriented in a compatible way with  the   time-orientation of  the hypersurface,  crosses $\cS$  in just one of the two possible   directions, the other being forbidden}. To the best of our knowledge,  such a    strong  constraint on   the  possible ways  of   crossing the null hypersurfaces  has not been   stated  so far in the literature, despite of the fact that it is quite  reasonable to expect  it, especially if one has in mind that the null hypersurfaces are  somehow intermediate between the time-like and  the space-like hypersurfaces, so that   in certain regards they are   similar to the former, in others    similar to the latter.   In order to  make fully clear what we mean,   we now  very  briefly review some basic facts  on time-like and space-like hypersurfaces. \par
\smallskip
We  recall  that imposing  a {\it time-orientation} on  a Lorentzian manifold corresponds to  selecting a  (smoothly varying)  family   $\sT = \{\sT_x\}_{x \in M}$  of connected components $\sT_x$  of  the cones $\sC_x \subset T_x M \setminus \{0\}$ of   causal (=  time-like or null)  non-zero vectors.   To make such a choice  one can just pick a  nowhere vanishing smooth causal vector field $Y$  and declare that, for any  $x \in N$,  the connected component  $\sT_x \subset \sC_x$ at $x$ of the time-orientation is the one  containing $Y_x$. There do exist Lorentzian manifolds on which  no global   time-orientation can be imposed, but on any sufficiently small open set  there are  always exactly  two  time-orientations which can be imposed, one opposite to the other.  \par
\smallskip
If $\cS$ is a {\it time-like hypersurface} in  a Lorentzian $n$-manifold $(M, g)$, then  $(\cS, g|_{T\cS \times T\cS})$ is a Lorentzian $(n-1)$-manifold and any time-orientation on $\cS$ (determined by a nowhere vanishing  causal vector field $Y$ in $T\cS$)  fixes  a  time-orientation on a neighbourhood $\cU \subset M$ of $\cS$. Indeed,  it suffices to extend  $Y$ to a  smooth vector field on $M$, so that one may  consider a  neighbourhood $\cU$ of $\cS$ where the extended $Y$ is still causal and non-zero and use $Y|_\cU$ to impose a time-orientation on $\cU$.  Nothing of this kind can be done if $\cS$ is a {\it space-like hypersurface}, because  no nowhere vanishing vector field $Y$ in  $T\cS$   is causal   and  no time-orientation on   neighbourhoods  can be   determined as above  (Fig.1 and Fig.2). \par
\begin{center}
\begin{tikzpicture}[scale=0.5]
\tikzstyle{every label}=[transparent]
		\node [style=every label] (0) at (-6, 4.5) {0};
		\node [style=every label] (1) at (-2, 6) {1};
		\node [style=every label] (2) at (8, 3.5) {2};
		\node [style=every label] (3) at (6.75, -1) {3};
		\node [style=every label] (4) at (-3.5, -2.75) {4};
		\node [style=every label] (5) at (-1, 5.5) {5};
		\node [style=every label] (6) at (-1.75, 5) {6};
		\node [style=every label] (7) at (1.5, 3.75) {7};
		\node [style=every label] (8) at (1.75, 3) {8};
		\node [style=every label] (9) at (2, -2) {9};
		\node [style=every label] (10) at (1.5, -2.5) {10};
		\node [style=every label] (11) at (-1, -0.5) {11};
		\node [style=every label] (12) at (-1.5, -0.25) {12};
		\node [style=every label] (13) at (0.5, 4.5) {13};
		\node [style=every label] (14) at (0.25, -1.5) {14};
		\node [style=every label] (20) at (4.25, 3.25) {20};
		\node [style=every label] (21) at (5.75, 0.75) {21};
		\node [style=every label] (22) at (4.25, 0.75) {22};
		\node [style=every label] (23) at (5.75, 3.25) {23};
		\node [style=every label] (24) at (5, 2) {24};
		\node [style=every label] (25) at (-4, 3.25) {25};
		\node [style=every label] (26) at (-5, 0.25) {26};
		\node [style=every label] (27) at (-6, 1) {27};
		\node [style=every label] (28) at (-3, 2.5) {28};
		\node [style=every label] (29) at (-4.5, 1.75) {29};
		\node [style=every label] (35) at (-1, 3.25) {35};
		\node [style=every label] (36) at (1, 0.25) {36};
		\node [style=every label] (37) at (-1, 0.25) {37};
		\node [style=every label] (38) at (1, 3.25) {38};
		\node [style=every label] (39) at (0, 1.75) {39};
		\node [style=every label] (40) at (-0.5, 3.5) {40};
		\node [style=every label] (41) at (0.5, 3) {41};
		\node [style=every label] (42) at (-0.5, 0.5) {42};
		\node [style=every label] (43) at (0.5, 0) {43};
		\node [style=every label] (44) at (0, 3.25) {44};
		\node [style=every label] (45) at (2, 4.5) {45};
		\node [black, above] (46) at (2.5, 4.5) {
		 };
		 %
		 	 \path[shade, shading= radial, inner color = black!50, outer color =white, opacity = 0.5]{[in=-180, out=30] (0.center) to (1.center)}
		{ [in=120, out=0, looseness=0.75] (1.center) to (2.center)} 
		{ [in=60, out=-60, looseness=1.25] (2.center) to (3.center)}
		{ [in=0, out=-120, looseness=0.75] (3.center) to (4.center)}
		{ [in=-150, out=-180] (4.center) to (0.center)}
		 (1.center) to (2.center) to (3.center) to (4.center) to (0.center);
 \path[shade, left color = red!100, right color = black!100, opacity = 1]
		(40.center) to (41.center)  
		{ [in=-45, out=-180, looseness=0.75] (41.center) to (38.center)} 
		{ [in=-165 , out=90, looseness=0.75] (38.center) to (40.center)} 
		(40.center) to (41.center) to (38.center)  
	to (40.center)
	(40.center) to (39.center) to (38.center) to (40.center)  ;
	 \path[shade, left color = red!30, right color = black!50, opacity = 1]
	(39.center) to (42.center) to (43.center) to (36.center) to (39.center) 
	{ [in=-90, out=0, looseness=0.75] (43.center) to (36.center)}
		{ [in=0, out=105, looseness=0.50] (36.center) to (42.center)}
		(41.center) to (39.center) (39.center) to (43.center); 
		   \path[shade, shading = radial,  inner color = cyan!70, outer color = cyan!20,  opacity = 0.5]
		(12.center) to (6.center) to (5.center)
		to (13.center) to (7.center) to (8.center) to (9.center) to (10.center)
		to (14.center) to (11.center) to (12.center) 
		 {[in=150, out=90, looseness=0.75] (6.center) to (5.center)}
		 {[blue, thick][in=165, out=-30] (5.center) to (13.center)}
		 {[in=150, out=-15, looseness=1.25] (13.center) to (7.center)}
		{[in=90, out=-15, looseness=0.75] (7.center) to (8.center)}
		(8.center) to (9.center)
		{[bend left=45] (9.center) to (10.center)}
		{[in=-15, out=165, looseness=0.75] (10.center) to (14.center)}
		{[in=0, out=165, looseness=0.75] (14.center) to (11.center)}
		{[in=-105, out=180, looseness=0.75] (11.center) to (12.center)}; 
\path[shade, left color = yellow!100, right color = black!100, opacity = 0.8]
		(40.center) to (41.center)  
		{ [in=-45, out=-180, looseness=0.75] (41.center) to (35.center)} 
		{ [in=-165 , out=90, looseness=0.75] (35.center) to (40.center)} 
		(40.center) to (41.center)  
		 to (35.center) to (40.center)
		 (35.center) to (39.center) to (41.center) to (35.center)
		 		 { [in=-45, out=-180, looseness=0.75] (41.center) to (35.center)}  ; 
		 \path[shade, left color = yellow!30, right color = black!50, opacity = 0.8]
		 (39.center) to (42.center) to (43.center) to (39.center)
		 {[in=90, out=-165, looseness=0.75] (42.center) to (37.center)} 
		 { [in=-180, out=-90, looseness=0.50] (37.center) to (43.center)}
		 (42.center) to (37.center) to (43.center) to (42.center)
		 (39.center) to (37.center) to (43.center) to (39.center); 
	\path[shade, left color = red!100, right color = black!100, opacity = 0.5]
		{[bend right=60, looseness=0.50] (28.center) to (25.center)}
		(25.center) to (29.center) to (28.center) to cycle; 
		\path[shade, left color = yellow!100, right color = black!100, opacity = 0.8]
		(25.center) to (29.center) to (28.center) to cycle
		{[bend left=60, looseness=0.50] (28.center) to (25.center)}; 
\path[shade, left color = red!10, right color = black!30, opacity = 0.8]
	 (29.center) to (27.center) to (26.center) to (29.center)
{ [bend left=60, looseness=0.50] (27.center) to (26.center)};
\path[shade, left color = yellow!30, right color = black!50, opacity = 0.4]
	 (29.center) to (27.center) to (26.center) to (29.center)
{ [bend right=60, looseness=0.50] (27.center) to (26.center)};
\path[shade, left color = red!100, right color = black!100, opacity = 0.5]
		(24.center) to (23.center) to (20.center) to cycle
		{[bend right=60, looseness=0.50] (23.center) to (20.center)}; 
		\path[shade, left color = yellow!100, right color = black!100, opacity = 0.8]
		{ [bend left=60, looseness=0.50] (23.center) to (20.center)}
		(24.center) to (23.center) to (20.center) to cycle; 
\path[shade, left color = red!10, right color = black!30, opacity = 0.8]
	 (24.center) to (22.center) to (21.center) to cycle
{ [bend left=60, looseness=0.50] (22.center) to (21.center)};
\path[shade, left color = yellow!30, right color = black!50, opacity = 0.4]
	 (24.center) to (22.center) to (21.center) to cycle
{ [bend right=60, looseness=0.50] (22.center) to (21.center)};
\node [blue] (14*) at (1.25, 1.5) {\LARGE{$\cS$}};
\node [blue] (14**) at (2.1, 1.5) { \tiny\hskip 1 cm  time-like};
\node [black] (45*) at (4.8, 4.7) {$\sT^{(\cS)}_x = \sT_x \cap T_x \cS$
};
		\draw[thin, blue] (40.center) to (41.center);
		\draw[thin, blue] (42.center) to (43.center);
		\draw[thin, blue] (42.center) to (41.center);
		\draw[thin, blue] (40.center) to (43.center);
		\draw [thick, bend right, dashed] (45.center) to (44.center)[->];
\end{tikzpicture}
\end{center}
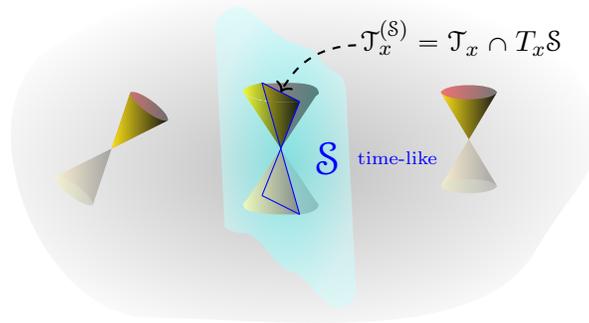
\centerline{\small{\bf Fig.1 --  Time-orientation  nearby a time-oriented time-like hypersurface}}
\begin{center}
    \begin{tikzpicture}[scale = 0.5]
      \tikzstyle{every label}=[transparent]
		\node [style=every label] (0) at (-6, 4.5) {0};
		\node [style=every label] (1) at (-2, 6) {1};
		\node [style=every label] (2) at (8, 3.5) {2};
		\node [style=every label] (3) at (6.75, -1) {3};
		\node [style=every label] (4) at (-3.5, -2.75) {4};
		\node [style=every label] (5) at (2.5, 1.75) {5};
		\node [style=every label] (6) at (2.75, 2.5) {6};
		\node [style=every label] (8) at (2, 1) {8};
		\node [style=every label] (9) at (-3, 1) {9};
		\node [style=every label] (20) at (4.25, 3.25) {20};
		\node [style=every label] (21) at (5.75, 0.75) {21};
		\node [style=every label] (22) at (4.25, 0.75) {22};
		\node [style=every label] (23) at (5.75, 3.25) {23};
		\node [style=every label] (24) at (5, 2) {24};
		\node [style=every label] (25) at (-4, 3.25) {25};
		\node [style=every label] (26) at (-5, 0.25) {26};
		\node [style=every label] (27) at (-6, 1) {27};
		\node [style=every label] (28) at (-3, 2.5) {28};
		\node [style=every label] (29) at (-4.5, 1.75) {29};
		\node [style=every label] (35) at (-1, 3.25) {35};
		\node [style=every label] (36) at (1, 0.25) {36};
		\node [style=every label] (37) at (-1, 0.25) {37};
		\node [style=every label] (38) at (1, 3.25) {38};
		\node [style=every label] (39) at (0, 1.75) {39};
		\node [style=every label] (40) at (-2.25, 2.5) {40};
		\node [style=every label] (41) at (-2.75, 1.75) {41};
		\node [style=every label] (42) at (0, 1.25) {42};
		\node [style=every label] (43) at (-0.5, 2.5) {43};
		\node [style=every label] (44) at (0.5, 2.5) {44};
		 \node [style=every label] (45) at (-0.5, 1) {45};
		\node [style=every label] (46) at (0.5, 1) {46};
		 	 \path[shade, shading= radial, inner color = black!50, outer color =white, opacity = 0.5]{[in=-180, out=30] (0.center) to (1.center)}
		{ [in=120, out=0, looseness=0.75] (1.center) to (2.center)} 
		{ [in=60, out=-60, looseness=1.25] (2.center) to (3.center)}
		{ [in=0, out=-120, looseness=0.75] (3.center) to (4.center)}
		{ [in=-150, out=-180] (4.center) to (0.center)}
		 (1.center) to (2.center) to (3.center) to (4.center) to (0.center);
 \path[shade, left color = red!10, right color = black!30, opacity = 0.5]
		(39.center) to (36.center)  to (37.center) to cycle
		{ [bend right=105, looseness=0.50] (36.center) to (37.center)};
 \path[shade, left color = yellow!10, right color = black!50, opacity = 0.8]
		(39.center) to (36.center)  to (37.center) to cycle
		{ [bend left=105, looseness=0.50] (36.center) to (37.center)};		
		  \path[shade, shading = radial,  inner color = cyan!70, outer color = cyan!20,  opacity = 0.8]
(40.center) to (43.center) to (44.center) to (6.center) to (5.center) to (8.center) to (46.center) to (45.center) to (9.center) to (41.center) to cycle 
{ [in=-75, out=0] (8.center) to (5.center)}
{ [in=-45, out=120, looseness=0.75] (5.center) to (6.center)}
{[in=165, out=-15, looseness=0.75] (44.center) to (6.center)}
{ [in=165, out=0] (40.center) to (43.center)}
{ [in=15, out=-180, looseness=1.25] (40.center) to (41.center)}
{[in=-180, out=-150] (41.center) to (9.center)}
{ [in=-165, out=0] (9.center) to (45.center)}
{[in=-165, out=15, looseness=1.25] (45.center) to (46.center)}
{ [in=180, out=15] (46.center) to (8.center)}
{ [in=-75, out=0] (8.center) to (5.center)}
{ [in=-45, out=120, looseness=0.75] (5.center) to (6.center)}
{ [in=165, out=-15, looseness=0.75] (44.center) to (6.center)};
 \path[shade, left color = red!100, right color = black!100, opacity = 0.5]
		(38.center) to (35.center)  to (39.center) to cycle
		{ [bend right=105, looseness=0.50] (38.center) to (35.center)};
 \path[shade, left color = yellow!100, right color = black!100, opacity = 0.8]
		(38.center) to (35.center)  to (39.center) to cycle
		{ [bend left=105, looseness=0.50] (38.center) to (35.center)};		
	\path[shade, left color = red!100, right color = black!100, opacity = 0.5]
		{[bend right=60, looseness=0.50] (28.center) to (25.center)}
		(25.center) to (29.center) to (28.center) to cycle; 
		\path[shade, left color = yellow!100, right color = black!100, opacity = 0.8]
		(25.center) to (29.center) to (28.center) to cycle
		{[bend left=60, looseness=0.50] (28.center) to (25.center)}; 
\path[shade, left color = red!10, right color = black!30, opacity = 0.8]
	 (29.center) to (27.center) to (26.center) to (29.center)
{ [bend left=60, looseness=0.50] (27.center) to (26.center)};
\path[shade, left color = yellow!30, right color = black!50, opacity = 0.4]
	 (29.center) to (27.center) to (26.center) to (29.center)
{ [bend right=60, looseness=0.50] (27.center) to (26.center)};
\path[shade, left color = red!100, right color = black!100, opacity = 0.5]
		(24.center) to (23.center) to (20.center) to cycle
		{[bend right=60, looseness=0.50] (23.center) to (20.center)}; 
		\path[shade, left color = yellow!100, right color = black!100, opacity = 0.8]
		{ [bend left=60, looseness=0.50] (23.center) to (20.center)}
		(24.center) to (23.center) to (20.center) to cycle; 
\path[shade, left color = red!10, right color = black!30, opacity = 0.8]
	 (24.center) to (22.center) to (21.center) to cycle
{ [bend left=60, looseness=0.50] (22.center) to (21.center)};
\path[shade, left color = yellow!30, right color = black!50, opacity = 0.4]
	 (24.center) to (22.center) to (21.center) to cycle
{ [bend right=60, looseness=0.50] (22.center) to (21.center)};
\node [blue] (14*) at (1.25, 1.5) {\LARGE{$\cS$}};
\node [blue] (14**) at (2.1, 1.5) { \tiny\ \ \ \hskip 0.5 cm  space-like};
\end{tikzpicture}
\end{center}
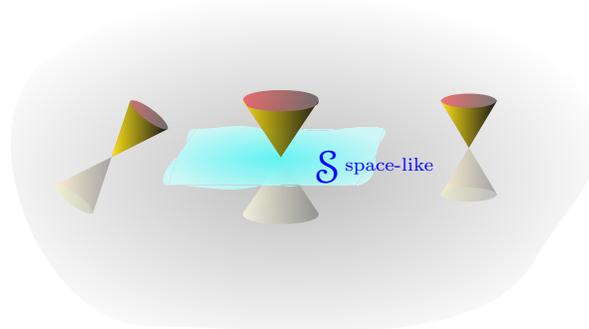
\centerline{\small{\bf  Fig.2 --  One of the two  freely choosable  time-orientations}}
\centerline{\small{\bf    on a neighbourhood of  a space-like hypersurface }}
\ \par
A similar contrast occurs  in the  directions  of crossing a hypersurface by  causal world-lines. Indeed,  if $\cS$ is a time-oriented, time-like hypersurface  and   $\sT = \{\sT_x\}_{x \in \cU}$ is the associated  time-orientation of a neighbourhood $\cU$ of $\cS$,  for any $x_o \in \cS$ there  exist  causal world-lines, which are time-oriented in a compatible  way with  $\sT  $  and cross  $\cS$ at $x_o$   in either  one of the two possible  ways of crossing.  A simple way to check this   consists  in  selecting a time-like vector $v_o \in \sT_{x_o} \cap  T_{x_o}\cS$, a space-like vector $w_o$ in $ T_{x_o} M \setminus T_{x_o} \cS$  and  two  perturbations of $v_o$ of the form  $v^{(\ve)}_\pm = v_o \pm \ve w_o  \in T_{x_o} M \setminus T_{x_o} \cS$. If $\ve$ is sufficiently small,  both perturbations are  in    $\sT_{x_o} \setminus T_{x_o} \cS$ and the  traces of the time-like geodesics,   passing through $x_o$ and tangent to $v^{(\ve)}_{+}$ and $v^{(\ve)}_-$, respectively,  are   causal world-lines  that are time-oriented in a compatible way  and having  the following property: one  crosses $\cS$ in one  direction,   the other in the opposite  (Fig.3). 
 Consider now a point  $x_o$  of a space-like hypersurface $\cS \subset M$ and pick  a  time-orientation   $\sT = \{\sT_x\}_{x \in \cU}$  on a neighbourhood $\cU$ of $\cS$ (it does not matter which of the two possible ones). In this situation there is only one possible way for a $\sT$-oriented causal world-line to cross $\cS$ at $x_o$:  All tangent vectors at $x_o$  of  such time-oriented world-lines have  to be in $\sT_{x_o}$ and hence all of them have  to   point  towards the same side of $\cS$ (Fig.4). \par
\smallskip
\begin{center}
    \begin{tikzpicture}[scale = 0.5]
      \tikzstyle{every label}=[transparent]
        \tikzstyle{new label}=[transparent]
		\node [style=every label] (0) at (-6, 4.5) {0};
		\node [style=every label] (1) at (-2, 6) {1};
		\node [style=every label] (2) at (8, 3.5) {2};
		\node [style=every label] (3) at (6.75, -1) {3};
		\node [style=every label] (4) at (-3.5, -2.75) {4};
		\node [style=every label] (5) at (-1, 5.5) {5};
		\node [style=every label] (6) at (-1.75, 5) {6};
		\node [style=every label] (7) at (1.5, 3.75) {7};
		\node [style=every label] (8) at (1.75, 3.25) {8};
		\node [style=every label] (9) at (2, -2) {9};
		\node [style=every label] (10) at (1.5, -2.5) {10};
		\node [style=every label] (11) at (-1, -0.5) {11};
		\node [style=every label] (12) at (-1.5, -0.25) {12};
		\node [style=every label] (13) at (0.5, 4.5) {13};
		\node [style=every label] (14) at (0.25, -1.5) {14};
		\node [style=every label] (20) at (4.25, 3.25) {20};
		\node [style=every label] (21) at (5.75, 0.75) {21};
		\node [style=every label] (22) at (4.25, 0.75) {22};
		\node [style=every label] (23) at (5.75, 3.25) {23};
		\node [style=every label] (24) at (5, 2) {24};
		\node [style=every label] (25) at (-4, 3.25) {25};
		\node [style=every label] (26) at (-5, 0.25) {26};
		\node [style=every label] (27) at (-6, 1) {27};
		\node [style=every label] (28) at (-3, 2.5) {28};
		\node [style=every label] (29) at (-4.5, 1.75) {29};
		\node [style=every label] (35) at (-1, 3.25) {35};
		\node [style=every label] (36) at (1, 0.25) {36};
		\node [style=every label] (37) at (-1, 0.25) {37};
		\node [style=every label] (38) at (1, 3.25) {38};
		\node [style=every label] (39) at (0, 1.75) {39};
		\node [style=every label] (40) at (-0.5, 3.5) {40};
		\node [style=every label] (41) at (0.5, 3) {41};
		\node [style=every label] (42) at (-0.5, 0.5) {42};
		\node [style=every label] (43) at (0.5, 0) {43};
		\node [style=every label] (44) at (0, 3.25) {44};
		\node [style=every label] (45) at (1.5, 5) {45};
		\node [style=every label] (46) at (2, 5) {46};
		\node [style=new label] (47) at (-0.5, 2.75) {47};
		\node [style=new label] (48) at (0.25, 2.75) {48};
		\node [style=new label] (49) at (-4.25, 5) {49};
		\node [style=new label] (50) at (4.25, 0) {50};
		\node [style=new label] (51) at (1.75, 0.25) {51};
		\node [style=new label] (52) at (-4.5, -1) {52};
		\node [style=new label] (53) at (4, 5.25) {53};
		 	 \path[shade, shading= radial, inner color = black!50, outer color =white, opacity = 0.5]{[in=-180, out=30] (0.center) to (1.center)}
		{ [in=120, out=0, looseness=0.75] (1.center) to (2.center)} 
		{ [in=60, out=-60, looseness=1.25] (2.center) to (3.center)}
		{ [in=0, out=-120, looseness=0.75] (3.center) to (4.center)}
		{ [in=-150, out=-180] (4.center) to (0.center)}
		 (1.center) to (2.center) to (3.center) to (4.center) to (0.center);
		 
		 
 \path[shade, left color = red!100, right color = black!100, opacity = 1]
		(40.center) to (41.center)  
		{ [in=-45, out=-180, looseness=0.75] (41.center) to (38.center)} 
		{ [in=-165 , out=90, looseness=0.75] (38.center) to (40.center)} 
		(40.center) to (41.center) to (38.center)  
	to (40.center)
	(40.center) to (39.center) to (38.center) to (40.center)  ;
	 \path[shade, left color = red!30, right color = black!50, opacity = 1]
	(39.center) to (42.center) to (43.center) to (36.center) to (39.center) 
	{ [in=-90, out=0, looseness=0.75] (43.center) to (36.center)}
		{ [in=0, out=105, looseness=0.50] (36.center) to (42.center)}
		(41.center) to (39.center) (39.center) to (43.center); 
		 
		   \path[shade, shading = radial,  inner color = cyan!70, outer color = cyan!20,  opacity = 0.5]
		(12.center) to (6.center) to (5.center)
		to (13.center) to (7.center) to (8.center) to (9.center) to (10.center)
		to (14.center) to (11.center) to (12.center) 
		 {[in=150, out=90, looseness=0.75] (6.center) to (5.center)}
		 {[blue, thick][in=165, out=-30] (5.center) to (13.center)}
		 {[in=150, out=-15, looseness=1.25] (13.center) to (7.center)}
		{[in=90, out=-15, looseness=0.75] (7.center) to (8.center)}
		(8.center) to (9.center)
		{[bend left=45] (9.center) to (10.center)}
		{[in=-15, out=165, looseness=0.75] (10.center) to (14.center)}
		{[in=0, out=165, looseness=0.75] (14.center) to (11.center)}
		{[in=-105, out=180, looseness=0.75] (11.center) to (12.center)}; 


\draw[very thick, ->]  (39.center) to (48.center);

\path[shade, left color = yellow!100, right color = black!100, opacity = 0.8]
		(40.center) to (41.center)  
		{ [in=-45, out=-180, looseness=0.75] (41.center) to (35.center)} 
		{ [in=-165 , out=90, looseness=0.75] (35.center) to (40.center)} 
		(40.center) to (41.center)  
		 to (35.center) to (40.center)
		 (35.center) to (39.center) to (41.center) to (35.center)
		 		 { [in=-45, out=-180, looseness=0.75] (41.center) to (35.center)}  ; 
		 \path[shade, left color = yellow!30, right color = black!50, opacity = 0.8]
		 (39.center) to (42.center) to (43.center) to (39.center)
		 {[in=90, out=-165, looseness=0.75] (42.center) to (37.center)} 
		 { [in=-180, out=-90, looseness=0.50] (37.center) to (43.center)}
		 (42.center) to (37.center) to (43.center) to (42.center)
		 (39.center) to (37.center) to (43.center) to (39.center); 
		 

	\path[shade, left color = red!100, right color = black!100, opacity = 0.5]
		{[bend right=60, looseness=0.50] (28.center) to (25.center)}
		(25.center) to (29.center) to (28.center) to cycle; 
		\path[shade, left color = yellow!100, right color = black!100, opacity = 0.8]
		(25.center) to (29.center) to (28.center) to cycle
		{[bend left=60, looseness=0.50] (28.center) to (25.center)}; 
\path[shade, left color = red!10, right color = black!30, opacity = 0.8]
	 (29.center) to (27.center) to (26.center) to (29.center)
{ [bend left=60, looseness=0.50] (27.center) to (26.center)};
\path[shade, left color = yellow!30, right color = black!50, opacity = 0.4]
	 (29.center) to (27.center) to (26.center) to (29.center)
{ [bend right=60, looseness=0.50] (27.center) to (26.center)};

\path[shade, left color = red!100, right color = black!100, opacity = 0.5]
		(24.center) to (23.center) to (20.center) to cycle
		{[bend right=60, looseness=0.50] (23.center) to (20.center)}; 
		\path[shade, left color = yellow!100, right color = black!100, opacity = 0.8]
		{ [bend left=60, looseness=0.50] (23.center) to (20.center)}
		(24.center) to (23.center) to (20.center) to cycle; 
\path[shade, left color = red!10, right color = black!30, opacity = 0.8]
	 (24.center) to (22.center) to (21.center) to cycle
{ [bend left=60, looseness=0.50] (22.center) to (21.center)};
\path[shade, left color = yellow!30, right color = black!50, opacity = 0.4]
	 (24.center) to (22.center) to (21.center) to cycle
{ [bend right=60, looseness=0.50] (22.center) to (21.center)};

		\draw[thin, blue] (40.center) to (41.center);
		\draw[thin, blue] (42.center) to (43.center);
		\draw[thin, blue] (42.center) to (41.center);
		\draw[thin, blue] (40.center) to (43.center);

		\draw[thick, ->] (39.center) to (47.center);
		
		\draw [in=315, out=120, red] (39.center) to (49.center)[->];
		\draw [bend left=15, red] (50.center) to (51.center)[->];
		\draw [bend left=15, red, dashed] (51.center) to (39.center)[->];
		\draw [bend right, looseness=0.75, red] (52.center) to (39.center);
		\draw [in=195, out=15, looseness=0.75, red] (7.center) to (53.center)[->];
		\draw [in=75, out=-165, looseness=0.75, dashed, red] (7.center) to (39.center)[->];
\end{tikzpicture}
\end{center}
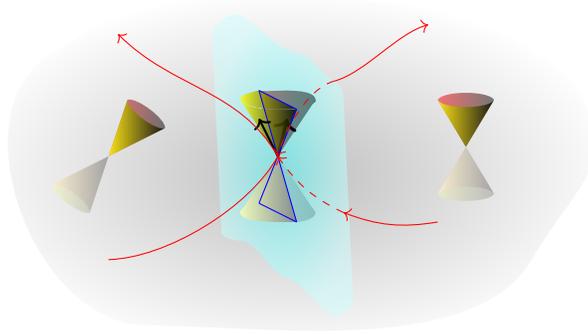
\centerline{\small{\bf Fig.3 --  The two ways of crossing a time-oriented time-like hypersurface}}
\begin{center}
    \begin{tikzpicture}[scale = 0.5]
      \tikzstyle{every label}=[transparent]
        \tikzstyle{new label}=[transparent]
		\node [style=every label] (0) at (-6, 4.5) {0};
		\node [style=every label] (1) at (-2, 6) {1};
		\node [style=every label] (2) at (8, 3.5) {2};
		\node [style=every label] (3) at (6.75, -1) {3};
		\node [style=every label] (4) at (-3.5, -2.75) {4};
		\node [style=every label] (5) at (2.5, 1.75) {5};
		\node [style=every label] (6) at (2.75, 2.5) {6};
		\node [style=every label] (8) at (2, 1) {8};
		\node [style=every label] (9) at (-3, 1) {9};
		\node [style=every label] (20) at (4.25, 3.25) {20};
		\node [style=every label] (21) at (5.75, 0.75) {21};
		\node [style=every label] (22) at (4.25, 0.75) {22};
		\node [style=every label] (23) at (5.75, 3.25) {23};
		\node [style=every label] (24) at (5, 2) {24};
		\node [style=every label] (25) at (-4, 3.25) {25};
		\node [style=every label] (26) at (-5, 0.25) {26};
		\node [style=every label] (27) at (-6, 1) {27};
		\node [style=every label] (28) at (-3, 2.5) {28};
		\node [style=every label] (29) at (-4.5, 1.75) {29};
		\node [style=every label] (35) at (-1, 3.25) {35};
		\node [style=every label] (36) at (1, 0.25) {36};
		\node [style=every label] (37) at (-1, 0.25) {37};
		\node [style=every label] (38) at (1, 3.25) {38};
		\node [style=every label] (39) at (0, 1.75) {39};
		\node [style=every label] (40) at (-2.25, 2.5) {40};
		\node [style=every label] (41) at (-2.75, 1.75) {41};
		\node [style=every label] (42) at (0, 1.25) {42};
		\node [style=every label] (43) at (-0.5, 2.5) {43};
		\node [style=every label] (44) at (0.5, 2.5) {44};
		\node [style=new label] (45) at (0.25, 2.75) {45};
		\node [style=new label] (46) at (-0.25, 2.75) {46};
		\node [style=new label] (47) at (-2.25, -1) {47};
		\node [style=new label] (48) at (-2.5, 4.75) {48};
		\node [style=new label] (49) at (1.75, -1.25) {49};
		\node [style=new label] (50) at (2.25, 5) {50};
		\node [style=new label] (51) at (-0.25, 1.25) {51};
		\node [style=new label] (52) at (0.25, 1.25) {52};

		 	 \path[shade, shading= radial, inner color = black!50, outer color =white, opacity = 0.5]{[in=-180, out=30] (0.center) to (1.center)}
		{ [in=120, out=0, looseness=0.75] (1.center) to (2.center)} 
		{ [in=60, out=-60, looseness=1.25] (2.center) to (3.center)}
		{ [in=0, out=-120, looseness=0.75] (3.center) to (4.center)}
		{ [in=-150, out=-180] (4.center) to (0.center)}
		 (1.center) to (2.center) to (3.center) to (4.center) to (0.center);

 \path[shade, left color = red!10, right color = black!30, opacity = 0.5]
		(39.center) to (36.center)  to (37.center) to cycle
		{ [bend right=105, looseness=0.50] (36.center) to (37.center)};
 \path[shade, left color = yellow!10, right color = black!50, opacity = 0.8]
		(39.center) to (36.center)  to (37.center) to cycle
		{ [bend left=105, looseness=0.50] (36.center) to (37.center)};
		

		  \path[shade, shading = radial,  inner color = cyan!70, outer color = cyan!20,  opacity = 0.8]
(40.center) to (43.center) to (44.center) to (6.center) to (5.center) to (8.center) to (42.center) to (9.center) to (41.center) to cycle 
{ [in=-75, out=0] (8.center) to (5.center)}
{ [in=-45, out=120, looseness=0.75] (5.center) to (6.center)}
{[in=165, out=-15, looseness=0.75] (44.center) to (6.center)}
{ [in=165, out=0] (40.center) to (43.center)}
{ [in=15, out=-180, looseness=1.25] (40.center) to (41.center)}
{[in=-180, out=-150] (41.center) to (9.center)}
{ [in=-165, out=0] (9.center) to (42.center)}
{ [in=180, out=15] (42.center) to (8.center)}
{ [in=-75, out=0] (8.center) to (5.center)}
{ [in=-45, out=120, looseness=0.75] (5.center) to (6.center)}
{ [in=165, out=-15, looseness=0.75] (44.center) to (6.center)};

	;


 \path[shade, left color = red!100, right color = black!100, opacity = 0.5]
		(38.center) to (35.center)  to (39.center) to cycle
		{ [bend right=105, looseness=0.50] (38.center) to (35.center)};
 \path[shade, left color = yellow!100, right color = black!100, opacity = 0.8]
		(38.center) to (35.center)  to (39.center) to cycle
		{ [bend left=105, looseness=0.50] (38.center) to (35.center)};
		

	\path[shade, left color = red!100, right color = black!100, opacity = 0.5]
		{[bend right=60, looseness=0.50] (28.center) to (25.center)}
		(25.center) to (29.center) to (28.center) to cycle; 
		\path[shade, left color = yellow!100, right color = black!100, opacity = 0.8]
		(25.center) to (29.center) to (28.center) to cycle
		{[bend left=60, looseness=0.50] (28.center) to (25.center)}; 
\path[shade, left color = red!10, right color = black!30, opacity = 0.8]
	 (29.center) to (27.center) to (26.center) to (29.center)
{ [bend left=60, looseness=0.50] (27.center) to (26.center)};
\path[shade, left color = yellow!30, right color = black!50, opacity = 0.4]
	 (29.center) to (27.center) to (26.center) to (29.center)
{ [bend right=60, looseness=0.50] (27.center) to (26.center)};
		
\path[shade, left color = red!100, right color = black!100, opacity = 0.5]
		(24.center) to (23.center) to (20.center) to cycle
		{[bend right=60, looseness=0.50] (23.center) to (20.center)}; 
		\path[shade, left color = yellow!100, right color = black!100, opacity = 0.8]
		{ [bend left=60, looseness=0.50] (23.center) to (20.center)}
		(24.center) to (23.center) to (20.center) to cycle; 
\path[shade, left color = red!10, right color = black!30, opacity = 0.8]
	 (24.center) to (22.center) to (21.center) to cycle
{ [bend left=60, looseness=0.50] (22.center) to (21.center)};
\path[shade, left color = yellow!30, right color = black!50, opacity = 0.4]
	 (24.center) to (22.center) to (21.center) to cycle
{ [bend right=60, looseness=0.50] (22.center) to (21.center)};

\draw[very thick, blue] (39.center) to (0.4,3.3)[->];
		\draw[very thick, blue] (39.center) to (-0.4,3.4)[->];
		\draw [bend left=15, red] (47.center) to (51.center)[->];
		\draw [bend right=15, red, dashed] (51.center) to (39.center);
		\draw [bend right=15, red] (49.center) to (52.center)[->];
		\draw [bend left=15, red, dashed] (52.center) to (39.center);
		\draw [in=-30, out=105, red] (39.center) to (48.center)[->];
		\draw [in=-135, out=75, red] (39.center) to (50.center)[->];
\end{tikzpicture}
\end{center}
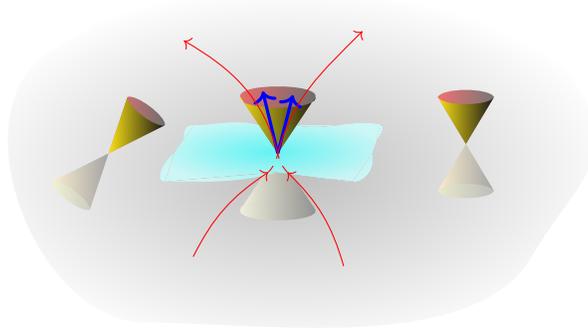
\centerline{\small{\bf  Fig.4 -- The only way of crossing a space-like hypersurface}}
\centerline{\small{\bf(provided that  a nearby  time-orientation  is  prescribed)}}
\ \par
The {\it null} (or {\it light-like}) {\it  hypersurfaces} of    $(M, g)$ constitute a sort of borderland  between the classes of   time-like and  space-like hypersurfaces  -- indeed, any   smooth homotopy  of  hypersurfaces, which begins with  a space-like hypersurface and ends with a time-like one,  must  contain a  null hypersurface  at some intermediate step.  The   properties of null hypersurfaces have been intensively studied in the literatures and we refer the reader to   \cite{DB,  JKC} and references therein. Here we just need to recall that any null hypersurface  is  equipped with a rank one distribution $\cD \subset T\cS$ consisting of  light-like vectors.    For   any  nowhere vanishing  vector field $E_o$ in  $\cD$, we may 
consider the unique family $\sT^{(\cS)} = \{\sT_x\}_{x \in \cS}$ of  connected components $\sT_x$ of the causal cones $\sC_x  \subset T_x M \setminus \{0\}$, $x \in \cS$,   containing the light-like vectors $E_{ox}$.  By a small abuse of language, we call  $\sT^{(\cS)}$ the  {\it time-orientation} of $\cS$ determined by $E_o$. Note that any causal smooth  extension  of  $E_o$ to a neighbourhood $\cU$ of $\cS$   determines a  time-orientation $\sT = \{\sT_x\}_{x \in \cU}$ on  $\cU$, which is also the unique time-orientation  on  $\cU$  that   induces the family of connected components  $\sT^{(\cS)}$ at the points of $\cS$.  \par
Being intermediate  between the time-like and the space-like, it is   sensible to expect  that some aspects of  the null hypersurfaces  are  similar to those of   the time-like, while  others are  similar to those of the space-like.  This expectation is fully confirmed by the following two properties: (a) On the one hand, as  we just pointed out, {\it any time-orientation of a null hypersurface  $\cS$ admits a canonical extension to a time-like orientation of a neighbourhood $\cU$, exactly as it occurs for the time-like  and in contrast with the space-like} (Fig. 5);  (b) One the other hand, our main result states that  {\it any time-oriented  null hypersurface  can be crossed by  appropriately time-oriented world-lines in just one of the two conceivable ways, exactly as it occurs  for the space-like  and in opposition to the time-like}  (Fig. 6).
 \par
\ \\[-1.2cm]
 \begin{center}
    \begin{tikzpicture}[scale= 0.5]
    \tikzstyle{every label}=[transparent]
		\node [style=every label] (0) at (-6, 4.5) {0};
		\node [style=every label] (1) at (-2, 6) {1};
		\node [style=every label] (2) at (8, 3.5) {2};
		\node [style=every label] (3) at (6.75, -1) {3};
		\node [style=every label] (4) at (-3.5, -2.75) {4};
		\node [style=every label] (5) at (2, 5) {5};
		\node [style=every label] (6) at (2, 5.25) {6};
		\node [style=every label] (8) at (2.25, 4.5) {8};
		\node [style=every label] (9) at (-1.75, -1.75) {9};
		\node [style=every label] (11) at (-2, -1.5) {11};
		\node [style=every label] (12) at (-2.5, -1.25) {12};
		\node [style=every label] (20) at (4.25, 3.25) {20};
		\node [style=every label] (21) at (5.75, 0.75) {21};
		\node [style=every label] (22) at (4.25, 0.75) {22};
		\node [style=every label] (23) at (5.75, 3.25) {23};
		\node [style=every label] (24) at (5, 2) {24};
		\node [style=every label] (25) at (-4, 3.25) {25};
		\node [style=every label] (26) at (-5, 0.25) {26};
		\node [style=every label] (27) at (-6, 1) {27};
		\node [style=every label] (28) at (-3, 2.5) {28};
		\node [style=every label] (29) at (-4.5, 1.75) {29};
		\node [style=every label] (35) at (-1, 3.25) {35};
		\node [style=every label] (36) at (1, 0.25) {36};
		\node [style=every label] (37) at (-1, 0.25) {37};
		\node [style=every label] (38) at (1, 3.25) {38};
		\node [style=every label] (39) at (0, 1.75) {39};
		
		 	 \path[shade, shading= radial, inner color = black!50, outer color =white, opacity = 0.5]{[in=-180, out=30] (0.center) to (1.center)}
		{ [in=120, out=0, looseness=0.75] (1.center) to (2.center)} 
		{ [in=60, out=-60, looseness=1.25] (2.center) to (3.center)}
		{ [in=0, out=-120, looseness=0.75] (3.center) to (4.center)}
		{ [in=-150, out=-180] (4.center) to (0.center)}
		 (1.center) to (2.center) to (3.center) to (4.center) to (0.center);

 \path[shade, left color = red!10, right color = black!30, opacity = 0.5]
		(39.center) to (36.center)  to (37.center) to cycle
		{ [bend right=105, looseness=0.50] (36.center) to (37.center)};
 \path[shade, left color = yellow!10, right color = black!50, opacity = 0.8]
		(39.center) to (36.center)  to (37.center) to cycle
		{ [bend left=105, looseness=0.50] (36.center) to (37.center)};
		

		  \path[shade, shading = radial,  inner color = cyan!70, outer color = cyan!20,  opacity = 0.5]
(8.center) to (9.center) to (11.center) to (12.center) to (6.center) to (5.center) to cycle
	{ [in=150, out=90, looseness=0.75] (6.center) to (5.center)}
	{[in=-105, out=180, looseness=0.75] (11.center) to (12.center)}
{[bend left] (5.center) to (8.center)}
	{ [bend left] (11.center) to (9.center)};


 \path[shade, left color = red!100, right color = black!100, opacity = 0.5]
		(38.center) to (35.center)  to (39.center) to cycle
		{ [bend right=105, looseness=0.50] (38.center) to (35.center)};
 \path[shade, left color = yellow!100, right color = black!100, opacity = 0.8]
		(38.center) to (35.center)  to (39.center) to cycle
		{ [bend left=105, looseness=0.50] (38.center) to (35.center)};
		

	\path[shade, left color = red!100, right color = black!100, opacity = 0.5]
		{[bend right=60, looseness=0.50] (28.center) to (25.center)}
		(25.center) to (29.center) to (28.center) to cycle; 
		\path[shade, left color = yellow!100, right color = black!100, opacity = 0.8]
		(25.center) to (29.center) to (28.center) to cycle
		{[bend left=60, looseness=0.50] (28.center) to (25.center)}; 
\path[shade, left color = red!10, right color = black!30, opacity = 0.8]
	 (29.center) to (27.center) to (26.center) to (29.center)
{ [bend left=60, looseness=0.50] (27.center) to (26.center)};
\path[shade, left color = yellow!30, right color = black!50, opacity = 0.4]
	 (29.center) to (27.center) to (26.center) to (29.center)
{ [bend right=60, looseness=0.50] (27.center) to (26.center)};
		
\path[shade, left color = red!100, right color = black!100, opacity = 0.5]
		(24.center) to (23.center) to (20.center) to cycle
		{[bend right=60, looseness=0.50] (23.center) to (20.center)}; 
		\path[shade, left color = yellow!100, right color = black!100, opacity = 0.8]
		{ [bend left=60, looseness=0.50] (23.center) to (20.center)}
		(24.center) to (23.center) to (20.center) to cycle; 
\path[shade, left color = red!10, right color = black!30, opacity = 0.8]
	 (24.center) to (22.center) to (21.center) to cycle
{ [bend left=60, looseness=0.50] (22.center) to (21.center)};
\path[shade, left color = yellow!30, right color = black!50, opacity = 0.4]
	 (24.center) to (22.center) to (21.center) to cycle
{ [bend right=60, looseness=0.50] (22.center) to (21.center)};
		\draw[line width = 2 pt, blue]  (39.center) to (38.center);
	\draw[very  thick, blue, dashed] (39.center) to (37.center);

\draw [thick, bend left, dashed] (3, 4.5) to (0.8, 2.5)[->];

\node [black] (45*) at (3.5, 4.9) {$\sT^{(\cS)}_x = \sT_x \cap T_x \cS$};
		\end{tikzpicture}
\end{center}
\ \\[-1 cm]
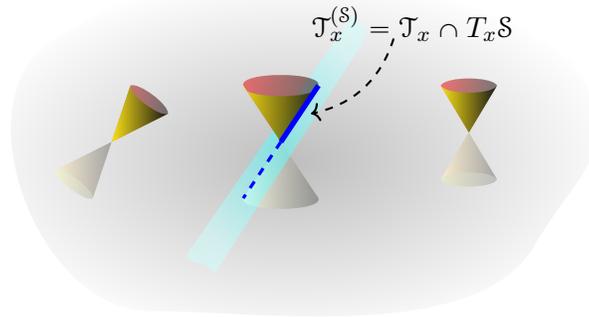
\centerline{\small{\bf Fig. 5 --  Time-orientation   nearby a time-oriented light-like hypersurface}}
\ \\[-1cm]
\begin{center}
    \begin{tikzpicture}[scale = 0.5]
      \tikzstyle{every label}=[transparent]
       \tikzstyle{new label}=[transparent]
		\node [style=every label] (0) at (-6, 4.5) {0};
		\node [style=every label] (1) at (-2, 6) {1};
		\node [style=every label] (2) at (8, 3.5) {2};
		\node [style=every label] (3) at (6.75, -1) {3};
		\node [style=every label] (4) at (-3.5, -2.75) {4};
		\node [style=every label] (5) at (2, 5) {5};
		\node [style=every label] (6) at (2, 5.25) {6};
		\node [style=every label] (8) at (2.25, 4.5) {8};
		\node [style=every label] (9) at (-1.75, -1.75) {9};
		\node [style=every label] (11) at (-2, -1.5) {11};
		\node [style=every label] (12) at (-2.5, -1.25) {12};
		\node [style=every label] (20) at (4.25, 3.25) {20};
		\node [style=every label] (21) at (5.75, 0.75) {21};
		\node [style=every label] (22) at (4.25, 0.75) {22};
		\node [style=every label] (23) at (5.75, 3.25) {23};
		\node [style=every label] (24) at (5, 2) {24};
		\node [style=every label] (25) at (-4, 3.25) {25};
		\node [style=every label] (26) at (-5, 0.25) {26};
		\node [style=every label] (27) at (-6, 1) {27};
		\node [style=every label] (28) at (-3, 2.5) {28};
		\node [style=every label] (29) at (-4.5, 1.75) {29};
		\node [style=every label] (35) at (-1, 3.25) {35};
		\node [style=every label] (36) at (1, 0.25) {36};
		\node [style=every label] (37) at (-1, 0.25) {37};
		\node [style=every label] (38) at (1, 3.25) {38};
		\node [style=every label] (39) at (0, 1.75) {39};
		\node [style=new label] (40) at (0.25, 2.75) {40};
		\node [style=new label] (41) at (-0.25, 2.75) {41};
		\node [style=new label] (42) at (-0.75, -2) {42};
		\node [style=new label] (43) at (0.75, -2) {43};
		\node [style=new label] (44) at (-0.25, 0.5) {44};
		\node [style=new label] (45) at (0, 1) {45};
		\node [style=new label] (46) at (-2.5, 5.25) {46};
		\node [style=new label] (47) at (1.75, 5.5) {47};
		
		 	 \path[shade, shading= radial, inner color = black!50, outer color =white, opacity = 0.5]{[in=-180, out=30] (0.center) to (1.center)}
		{ [in=120, out=0, looseness=0.75] (1.center) to (2.center)} 
		{ [in=60, out=-60, looseness=1.25] (2.center) to (3.center)}
		{ [in=0, out=-120, looseness=0.75] (3.center) to (4.center)}
		{ [in=-150, out=-180] (4.center) to (0.center)}
		 (1.center) to (2.center) to (3.center) to (4.center) to (0.center);

 \path[shade, left color = red!10, right color = black!30, opacity = 0.5]
		(39.center) to (36.center)  to (37.center) to cycle
		{ [bend right=105, looseness=0.50] (36.center) to (37.center)};
 \path[shade, left color = yellow!10, right color = black!50, opacity = 0.8]
		(39.center) to (36.center)  to (37.center) to cycle
		{ [bend left=105, looseness=0.50] (36.center) to (37.center)};
		

		  \path[shade, shading = radial,  inner color = cyan!70, outer color = cyan!20,  opacity = 0.5]
(8.center) to (9.center) to (11.center) to (12.center) to (6.center) to (5.center) to cycle
	{ [in=150, out=90, looseness=0.75] (6.center) to (5.center)}
	{[in=-105, out=180, looseness=0.75] (11.center) to (12.center)}
{[bend left] (5.center) to (8.center)}
	{ [bend left] (11.center) to (9.center)};


 \path[shade, left color = red!100, right color = black!100, opacity = 0.5]
		(38.center) to (35.center)  to (39.center) to cycle
		{ [bend right=105, looseness=0.50] (38.center) to (35.center)};
 \path[shade, left color = yellow!100, right color = black!100, opacity = 0.8]
		(38.center) to (35.center)  to (39.center) to cycle
		{ [bend left=105, looseness=0.50] (38.center) to (35.center)};
		

	\path[shade, left color = red!100, right color = black!100, opacity = 0.5]
		{[bend right=60, looseness=0.50] (28.center) to (25.center)}
		(25.center) to (29.center) to (28.center) to cycle; 
		\path[shade, left color = yellow!100, right color = black!100, opacity = 0.8]
		(25.center) to (29.center) to (28.center) to cycle
		{[bend left=60, looseness=0.50] (28.center) to (25.center)}; 
\path[shade, left color = red!10, right color = black!30, opacity = 0.8]
	 (29.center) to (27.center) to (26.center) to (29.center)
{ [bend left=60, looseness=0.50] (27.center) to (26.center)};
\path[shade, left color = yellow!30, right color = black!50, opacity = 0.4]
	 (29.center) to (27.center) to (26.center) to (29.center)
{ [bend right=60, looseness=0.50] (27.center) to (26.center)};
		
\path[shade, left color = red!100, right color = black!100, opacity = 0.5]
		(24.center) to (23.center) to (20.center) to cycle
		{[bend right=60, looseness=0.50] (23.center) to (20.center)}; 
		\path[shade, left color = yellow!100, right color = black!100, opacity = 0.8]
		{ [bend left=60, looseness=0.50] (23.center) to (20.center)}
		(24.center) to (23.center) to (20.center) to cycle; 
\path[shade, left color = red!10, right color = black!30, opacity = 0.8]
	 (24.center) to (22.center) to (21.center) to cycle
{ [bend left=60, looseness=0.50] (22.center) to (21.center)};
\path[shade, left color = yellow!30, right color = black!50, opacity = 0.4]
	 (24.center) to (22.center) to (21.center) to cycle
{ [bend right=60, looseness=0.50] (22.center) to (21.center)};
		\draw[line width = 3 pt, blue]  (39.center) to (38.center);
	\draw[very  thick, blue, dashed] (39.center) to (37.center);
		\draw[very thick, blue] (39.center) to (41.center)[->];
		\draw[very thick, blue] (39.center) to (40.center)[->];
		\draw [bend left=15, red] (42.center) to (44.center)[->];
		\draw [bend left=15, red] (43.center) to (45.center)[->];
		\draw [bend right=15, red, dashed] (44.center) to (39.center);
		\draw [in=-45, out=105, looseness=0.75, red]  (39.center) to (46.center)[->];
		\draw [red,  dashed] (45.center) to (39.center);
		\draw [red][in=-135, out=75, looseness=0.50] (39.center) to (47.center)[->];

\end{tikzpicture}
\end{center}
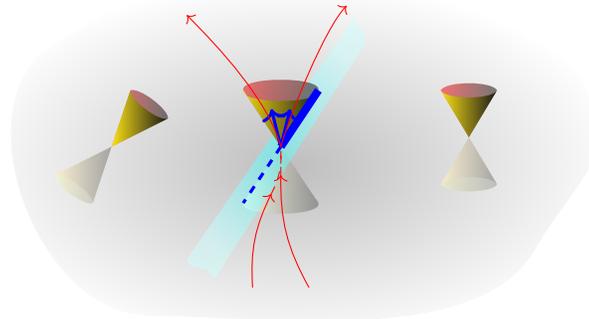
\centerline{\small{\bf  Fig. 6 -- The only way of crossing a time-oriented light-like hypersurface}}
\ \par
In order to get familiar with the relevance of   (b), consider the following  examples. A first and very easy one   is given  by  the upper  connected component of a  light cone in the Minkowski space $\bR^{1,3}$. Consider  cylindrical coordinates $((\xi, \psi ,r) ; t)$ for   $\bR^{1,3} \setminus \{x^1 = 0\}$, in which the  Minkowski metric takes the form
$ g_o = dt^2 -  dr^2 - r^2d\x^2-r^2\sin^2\x d\psi^2 $.
The   traces of the geodesics of $\bR^{1,3}$ are   segments and  the traces of the {\it  causal} geodesics are segments  parallel to  time-like or null vectors. The connected component    $\cS = \{  t  = r  \ , t > 0\}$ of the light cone $\cV = \{t = r\}$  is  a light-like  hypersurface  with  the following elementary property: 
{\it  No trace of a causal   geodesic of $\bR^{1,3}$ (= causal segment),  oriented towards   increasing values of the time $t$,    crosses $\cS$ starting  from  a point of the region  $ \{  t >  r \}$   and ending in the region  $\{t < r\}$}. This is because otherwise  the geodesic would be parallel to a space-like vector, hence not causal. A simple  argument easily  extends this ``no-crossing'' property to any other  causal world-line, not necessarily trace of a geodesic.\par
\medskip
A second (and much less trivial) illustration of  (b) is given by the  horizon of a Schwarz\-schild black hole. Consider the  manifold  $M_S= (\bR^3 \setminus \{0\})\times \bR $,  equipped with the Boyer-Lindquist coordinates  $((\xi, \psi ,r) ; u)$ and the  Schwarzschild metric 
 \beq\label{schwarzschild1}
g_S= \left (1- \frac{2m}{r}\right )du^2 -  2 du \vee dr - r^2d\x^2-r^2\sin^2\x d\psi^2 \ .
\eeq
 It is   well known that   $(M_S, g_S)$  has the
following  feature: {\it  No trace of a   causal geodesic,  oriented towards increasing values for the coordinate $u$,   starts from a point of the region   $  \{ r < 2m\}  $,   crosses the null hypersurface  $\cS = \{ r = 2m\} $ and ends  in the region $\{r > 2 m\}$.}  \par
\medskip
This    is the main aspect of the event horizons of all   known black holes metrics  (Schwarz\-schild, Reissner--Nordstr\"om, Kerr, Kerr-Newman, Myers-Perry, topological black holes, etc- -- see e.g.  \cite{AN, ABHP, BHTZ, BM, BLP, Ca0, Ca1, Ca2, Ca3, CN, DKS, MP, Va, KMV} and references therein). Even if it is known that any smooth event horizon (in the sense of Penrose, Hawking and Ellis)  has to be  a null hypersurface,   it is remarkable  that   the traditional way to verify whether a  null hypersurface has  the above semi-permeability property    is to   consider    explicit coordinate expressions for the equation of geodesics --as for  instance  using Boyer-Lindquist coordinates -- and to check  that no solution to the geodesic  equations can cross  the horizon in one of the two directions  (see  e.g. \cite{BL},  \cite[Appendix A]{Ca1}, \cite[Sect. 3.2]{MP}).  In contrast with this habit, our  result  unveils   that the  semi-permeability property  with respect to the causal geodesics  is  just a consequence  of the fact that  the horizons are   null hypersurfaces:  {\it No study of  the geodesic equations  is  needed to establish this property!}  Actually,  our result   {\it immediately} gives  the following   stronger property:   {\it  there is no   causal world-line {\rm (not necessarily trace of a  geodesic)},  which is appropriately time-oriented  and   crosses an event horizon $\cS$   from  the inner  region to the outer one} (\footnote{A direct proof of such  stronger   ``no-crossing property''   can be found in  \cite{Ch}, but only for  the   Schwarzschild  black holes. As far as we know,   a simple and explicit  proof   for  {\it any}  black hole was so far  missing.}). 
Motivated  by  all this,   we found convenient to consider  the next two  notions.\par
\medskip  
 Given a Lorentzian $n$-manifold  $(M, g)$, we say that {\it  a hypersurface $\cS \subset M$ is a semi-permeable barrier} (or  just {\it barrier}, for brevity) if (i) it is  time-oriented  and null  and (ii) it separates  $M$   into two regions, i.e.   $M \setminus \cS = M_+ \cup M_-$ with $M_\pm$  disjoint connected open sets,   both equipped with  a time-orientation which is compatible  with the time-orientation of $\cS$. By our  main result    {\it if $\cS$ is a barrier for $(M, g)$, then either  physical  signals can go from $M_+$ to $M_-$ but not  from  $M_-$ to $M_+$,  or vice versa}.    Note that  the  only allowed direction for the  crossing of  signals is completely  determined by  the   time-orientation of $\cS$ (and hence of  $M_\pm$ as well).  The allowed direction   is explicitly given  in   Theorem \ref{main} below. \par
 \medskip
Consider  now a  time-oriented region  (=connected open set) $\O_o$ of a Lorentzian $n$-manifold $(M, g)$ and assume that its  complement $\O^c_o = M \setminus \O_o$ consists of  a countable  union of  time-oriented regions $\O_1, \O_2, \ldots$ and of their   boundaries  $\cS_1 = \p \O_1$, $\cS_2 = \p \O_2$, $\ldots$ Assume also that  each $\cS_i$ is a time-oriented barrier and that the  following  holds:  for any index $i$, the time-orientations  of $\O_o$ and $\O_i$ are both compatible with the time-orientation of $\cS_i = \p \O_i$ and,   denoting by  $M_{(i)+}$, $M_{(i)-}$  the two  time-oriented components in which $M$ is separated by  $\cS_i$, then  either   $\O_i = M_{(i)+}$  and $\O_o \subset M_{(i)-}$, or  $\O_i = M_{(i)-}$  and $\O_o \subset M_{(i)+}$.  Under these condition,  we call a region  $\O_i \subset \O^c_o$   {\it black} (resp. {\it white}) {\it  for $\O_o$} if no physical signal can go from $\O_i$ to $\O_o$ (resp. from $\O_o$ to $\O_i$).  By the  properties of the barriers,  each   region $\O_i$ is either ``black'' or ``white'' for  $\O_o$.  \par
\medskip
 We now recall that  (see e.g. \cite{Pe, HE, Wl, Ca2, Ca3}), given  an asymptotically flat Lorentzian manifold $(M, g)$,  an {\it event horizon}  of $M$ is a  connected component of the boundary 
of a distinguished time-oriented region $\O_o \subset M$, named  {\it domain of outer communication} by Carter  in \cite{Ca2}.  It  essentially consists  of the  collection of  all points of $M$  that are causally joinable  to the infinite future boundary of the  observers. The {\it black holes} are  the connected components of the complement  $\O^c = M \setminus \overline \O_o$.   Note that, by a   lemma  of   Penrose on    achronal boundaries (\cite[Lemma I, p.\ 214]{Pe}), any event horizon is generated by segments of null geodesics so that,  in the cases in which it  is a smooth hypersurface,  such hypersurface is a  null one. 
Due to this, if the  domain of outer communication $\O_o \subset M$ has a smooth boundary, then each  connected component of its boundary is a barrier and the black holes  for $\O_o$  are black regions as we defined  above.  \par
\medskip
 In other words,  the class of  black (or possibly   white)  regions naturally include all black (or  white) holes with smooth boundaries.  It is however  important to keep in mind that the notion of  black/white region is strictly more general than what is usually understood by  the term ``(smoothly bounded) black/white hole''.  For instance,   the complement of a Schwarzschild black hole is a white region  for the inner part of the black hole, while  the observers that are located outside the black hole have never been described as  ``being sitting inside  a white hole'' (\footnote{Nonetheless,  such a term   might  be meaningful  if the space-time is considered under  the view point of an  observer  sitting {\it inside} a black hole. About  the possibility of the  existence  of  an observer of this kind, see the nice paper \cite{Do}}). Another  example of a black region, which cannot be taken as a black hole,   can be easily built considering   the half space $M_o = \{t> 0\}$ of the Minkowski space $\bR^{3,1}$, equipped with the  flat metric $ g_o$ and  the time-orientation corresponding to increasing  of  time-coordinate  $t$.  It suffices to observe  that, for such a  time-oriented Lorentzian manifold, the  (correspondingly time-oriented) half light-cone $\cS = \{  t  = r  \ , t > 0\}$ (\footnote{We recall that, in Penrose's terminology, this hypersurface  is a {\it particle horizon} \cite{Pe}.}) is a barrier  and the region $\O_1 = \{  t  > r \}$ is a black region for  $\O_o = M \setminus (\O_1 \cup \cS) = \{0 <  t< r\}$.   \par
 \medskip
  Due to the fact that the class of barriers  include all smooth event  horizons, the   task of locating the (smooth or piecewise smooth) event horizons for  space-times, equipped with numerically determined  metrics,  
  might be   reduced  to  locating  null hypersurfaces which separate the space-time.  In principle   this is a simpler problem than the previous. In fact,   determining event horizons requires to detect    {\it all} null geodesics  that  will neither  reach future null infinity  nor hit a singularity of the space-time (see e.g. \cite[\S\ 6.2]{BS} and references therein). This requires to  study systems of ode's that in principle might be quite involved.   The  problem of finding barriers is simpler, because a null hypersurface is    locally   a   level set $\{ F(x) = R\}$ for  a smooth function $F: \cU \subset M \to \bR$ with null differential at such level set.  This  condition  consists of the  single equation   $g^{-1}_{\{F = R\}}(d F, dF) = 0$,  which in general might be numerically  solved  either via integrations (see e.g. \cite[\S 2]{JKC}; see also \cite{ANN, LMSSW} for  interesting tricks to quickly integrate null hypersurfaces) or  minimising  the functional  $\mathcal F(F, R) = \int_{\cU} \big[g^{-1}_{\{F(x) = R\}}(d F, dF) \big]^2 \text{vol}_{g_o}$ for a Riemannian volume form   $\text{vol}_{g_o}$ in   the class of pairs, determined by   non-constant  functions $F: \cU \to \bR$ on  a fixed open set $\cU$ and   real constants $R\in \bR$. \par
  \medskip
We have to admit that  the  strategy  of determining event horizons by looking for them among the null hypersurfaces is not new --  it   has been frequently  used  in the literature.  For instance,  in  \cite{BL}  Boyer and Lindquist  describe the  event horizons as follows:  ``{in stationary space-times they  may  {\it provisionally} be defined as  stationary null hypersurfaces}''.   Much more recently, in \cite{GT},   G\"urses and Tekin  locate   dynamic horizons for a  Vaidya type dynamic black hole through  discussions of  certain types of null hypersurfaces.  However,  at the best of our knowledge,  such a strategy has been always   considered  as a sort of a  heuristic approach   to be  later confirmed  by     a posteriori  checks  of  either   the semi-permeability   feature   for  causal  geodesics or  the second law of black hole thermodynamics. By our  results,  no  further check   is needed to locate an event horizon (in the sense of Penrose, Hawking and Ellis) as soon as  one succeeds in  determining an outmost  barrier.\par 
\medskip
The structure of this paper is as follows.  In  \S \ref{preliminaries}  we review in great detail various definitions and properties concerning  orientations and time-orientations of hypersurfaces and curves in a Lorentzian manifold.  Then we  prove our main Theorem \ref{mainmain}  in \S \ref{mainsection}. In \S\ref{examples} we present   illustrative examples on how barriers and event horizons  can be determined using our Theorem \ref{mainmain}. For the benefit of the reader, in Appendix \ref{PHE}  we briefly outline the  classical  definitions of ``event horizon'' and ``black hole'', following  presentations of   a few  well known   textbooks.\par
\medskip
\section{Preliminaries on orientations,  time-orientations  and crossing}\label{preliminaries}
Here  we   review    basic definitions and  facts that  are exploited in  our main result. 
\par
\subsection{Oriented curves, oriented hypersurfaces and crossing} \label{crossing} Let $M$ be an $n$ dimensional manifold.  In this paper,  we use the 
term {\it open curve} to indicate a $1$-dimensional submanifold $\cC \subset M$, which is  homeomorphic to $\bR$. Due to this,   an open curve  always admits a global regular parameterisation (i.e.  
a smooth homeomorphism $\g:  (a, b)  \subset \bR\longrightarrow \cC \subset  M$ with  $\dot \g(s) \= \frac{d \g}{d s}\big|_s \neq 0$ for any $s \in  (a, b)$) (\footnote{In some  textbooks -- as e.g. in \cite{ON} -- our ``open curves''  are  called  ``world-lines''.  Here we call them differently   because    we are going to use the term  ``world-line''  with  a more restricted meaning.}).\par
\smallskip
Two global regular parameterisations $\g: (a, b) \to \cC$, $\g':  (\a, \b) \to \cC$ of the same open curve  are  called   {\it consonant} if  there is a smooth  change of parameter $h: (a,b) \longrightarrow (\a, \b)$ such that $\g(s) = \g'(h(s))$  with $\dot h(s) > 0$ at all $s$. The consonance  relation  is manifestly an equivalence relation  between   regular parameterisations  and there are  exactly two equivalence classes for such relation: If $\g: (a, b ) \to \cC$ is  in one class, the other class  is  the one of  $ \g(-t)$, $t\in (-b, -a)$.  
Each of these   equivalence classes  is named   {\it orientation} of $\cC$ and are denoted by $[\g]$ and $[-\g]$, respectively.  An {\it oriented  (open) curve} is an open curve $\cC \subset M$ equipped with an orientation. \par
\begin{rem} The above definition   of ``orientation of an open curve $\cC$'' is  perfectly equivalent to  the most   common  one,  namely to the  choice of   a smooth family  of non-zero  tangent vectors, each of them fixed  up to multiplication by a positive scalar. 
\end{rem}
A hypersurface $\cS \subset M$ is called {\it orientable} if  there exists at least one nowhere vanishing   vector field    $N: \cS \to TM|_{\cS}$ on $\cS$  with  
 $N_x \in  T_x M \setminus T_x \cS$ for any $x$. A vector field $N$ of this kind is called {\it   orienting vector  field for $\cS$}. Two orienting vector  fields $N$, $N'$ are said to be   {\it consonant} 
 if there is a smooth  positive  real function $\lm: \cS \to (0, + \infty)$ and a vector field $X: \cS \to TM|_\cS$, which is  tangent to $\cS$  at all points,  such that 
 $N' = \lm N + X$. As for the open curves,   such a  consonance relation   is an equivalence relation, with only  two equivalence classes per each orientable hypersurface $\cS$:  If  $N$ is an orienting vector field, one of the  two classes  is the equivalence class $[N]$ of   $N$, the other is the class  $[-N]$ containing $-N$. Each of these classes  is an   {\it orientation} of $\cS$. An  {\it oriented hypersurface} is an orientable hypersurface $\cS$ equipped with an orientation $[N]$.  
\par
 \smallskip
 \begin{rem}  As before,  this  notion of ``oriented hypersurface'' is  equivalent to all other  common   definitions, which  can be found  in   textbooks  on  Differential Geometry, as e.g. \cite{St,Wa}.  We adopt  this one simply because   it allows  shorter   proofs for  our    results. 
\end{rem}
\smallskip
Consider now  an open curve $\cC$ and  an orientable hypersurface $\cS$  of  $M$. We say that {\it $\cC$ intersects transversally $\cS$ at $x_o$} if $x_o \in \cC \cap \cS$ and $T_{x_o} \cC \cap T_{x_o} \cS = \{0\}$. This is equivalent to require that, given a global  regular parameterisation $\g:  (a, b) \to \cC$, if  $s_o \in (a, b)$  is such that $\g(s_o)  = x_o$,  then   $\dot \g(s_o) \notin T_{x_o} \cS$.  In case  $\cC$ is 
equipped with the orientation $[\g]$  and if $\cS$ is   equipped  with the orientation $[N]$,  we say that {\it the oriented curve $\cC$ crosses  $\cS$ at $x_o$ in the  direction of $[N]$} (resp. {\it of $[-N]$})  if 
$$ \dot \g(s_o) = \lm N_{x_o}  + v\  \qquad  \text{for some}\  \lm > 0\ \ (\text{resp.}\ \lm < 0\ )\ \  \text{and some}\ v \in T_{x_o} \cS\ . $$
 \par \medskip
\subsection{Causal world-lines and  time-orientations in Lorentzian manifolds}
Let $(M, g)$ be a Lorentzian manifold of dimension $n \geq 2$ with a  mostly minus signature. We recall that 
a   subspace $\{0\} \neq V \subset T_{x_o} M$ of a tangent space  $T_{x_o} M$ is called  {\it time-like} (resp. {\it light-like} or {\it null},   {\it space-like})   if  the restricted scalar product $g_{x_o}|_{V\times V}$  is  non-degenerate and such that $g_{x_o}(v, v) > 0$ for  some  $v \in V$,  thus Lorentzian whenever $\dim V \geq 2$
 (resp.  semi-negative degenerate,  negative definite). A non-zero  vector $0 \neq v \in T_{x_o} M$  is said to be  {\it time-like} (resp. {\it light-like} or {\it null},  {\it space-like})  if  it generates a  time-like (resp.  light-like,  space-like) one dimensional subspace. A vector $v \in T_{x_o} M$ is called {\it causal} if it is  either   time-like or null.\par
  According to  this terminology,  an open  curve $\cC$ (resp. a  hypersurface   $\cS$)   is  called {\it time-like}  (resp. {\it light-like} or {\it null}, {\it space-like}) if  its  tangent spaces  are all  time-like  (resp. null,   space-like). A curve is called {\it causal} if all of its non-zero tangent vectors are causal.
  \par
\smallskip
We recall  that if $\cS \subset M$ is a null hypersurface, then the collection of  its null tangent vectors  determine a distribution $\cD \subset T_x \cS$ of rank $1$. The integral leaves of this distribution are traces of geodesics and constitute  a geodesic congruence on $\cS$ (\cite{JKC}).\par
\smallskip
As we mentioned in the Introduction,  a {\it time-orientation}  of an open subset $\cU \subset M$ is a smooth  family  $\sT = \{\sT_x\}_{x \in \cU}$ of  connected components  (= half cones) $\sT_x \subset T_x M \setminus \{0\}$, $x \in \cU$,  of the cones of causal vectors   $\sC_x = \{v \in T_x M \setminus\{0\}\ :\  g_x(v,v) \geq  0\}$.   By  ``smooth family''   we mean that for any $x \in \cU$, there exists a  smooth  causal vector field $Y$ on  a neighbourhood  of $x$,  such that,   at all points  $y$ on which $Y$ is defined, the half-cone    $ \sT_y$  coincides with  the connected component of    $\sC_y$ to which $Y_y$ belongs.  The notions of time-orientability and  time-orientation immediately extend   to  time-like hypersurfaces, because they  are Lorentzian  if equipped with the induced metric.  As we already remarked,  they extend    to  null hypersurfaces  too, provided that the following  definition is adopted.
\begin{definition}\label{null-orientation} Let $\cS \subset M$ be   a null hypersurface and denote by $\cD \subset T \cS$ its $1$-distribution of null tangent vectors. We say that $\cS$ is {\it time-orientable} if there exists a nowhere vanishing vector field $E$ in $\cD$. In this case, we call {\it time-orientation for $\cS$ determined by $E$} the family  $\sT^{(\cS)} = \{\sT_x\}_{x \in \cS}$ of   the half-cones $\sT_x$ of the  causal cones $\sC_x  \subset T_x M \setminus \{0\}$, $x \in \cS$,   which contain the  vectors $E_{x}$. 
\end{definition}
Note that  if $\cS$ is null and time-orientable, then  it admits exactly two time-orientations. \par
\begin{rem}  We stress the fact  that  {\it a null hypersurface $\cS$ is time-orientable if and only if it is  orientable as a hypersurface}. Indeed, if there exists  a nowhere vanishing smooth vector field $E \in \cD$, at each  $x \in \cS$ we may consider a null vector $E^\perp_x \in T_x M \setminus T_x \cS$ that satisfies  $g(E_x, E^\perp_x) = 1$. Since this vector can be chosen smoothly depending on $x$,  this gives an orienting vector field $N = E^\perp$ for  the hypersurface $\cS$.  Conversely, if $\cS$ admits an orienting vector field $N \in TM|_{\cS} \setminus T \cS$, there exists a  space-like vector field $X \in T\cS$ such that $E^\perp = N - X$ is null (this can be  checked   working with the  expansion of  $N$  in terms of   locally defined  frame field  $(e_i)_{i = 1, \ldots, n}$  with vector fields $e_{3}, \ldots, e_{n}$  space-like and tangent to $\cS$ at the  points of the hypersurface). Thus also   $E^\perp|_{\cS}$  is an orienting vector field for $\cS$ and there is  a nowhere vanishing null  vector field $E$ in $\cD \subset T\cS$  with $g(E^\perp_x, E_x) = 1$ at any $x \in \cS$. This means that $\cS$ is time-orientable.
\end{rem}
We conclude this preliminary section with the next crucial definition.\par
\begin{definition} \label{admissible}   A  {\it causal world-line} (or just {\it world-line}, for short)  in an open set $\cU \subset M$ is a causal open curve $\cC \subset \cU$.  If  a world-line  $\cC$ is equipped with   an orientation $[\g]$ and if there is a time orientation $\sT$ on  an open subset $\cU' \subset \cU$  with  non-trivial intersection with  $\cC$,  we say that $\cC$ is {\it  $\sT$-time oriented} (resp. {\it  $-\sT$-time oriented})  if for one (and hence for any) point $x = \g(s) \in \cC \cap  \cU'$,  the non-zero    vector  $\dot \g(s)$  (resp. 
$- \dot \g(s)$)  is in  $ \sT_{x = \g(s)}$. 
\end{definition}
\par
\medskip
\section{The semi-permeability property of   null hypersurfaces}\label{mainsection}
 We start with  the following useful notion.  \par
\begin{definition} \label{def-barrier} Let $(M, g)$ be a Lorentzian $n$-manifold  with mostly negative signature.   A  {\it dressed  barrier  of  $M$} is a pair $(\cS, \bT)$ given by
\begin{itemize}
\item[--] a time-orientable null  hypersurface $\cS \subset M$  and
\item[--]  a smooth vector field $\bT$  on $M$
\end{itemize}
   satisfying the following   conditions:
\begin{itemize} [leftmargin = 18pt]
\item[(a)] $M \setminus \cS$ consists of exactly two  connected components, say $M_+$ and $M_-$; 
\item[(b)]  for any $y \in  \cS$ the vector $\bT_y$ is non-zero, null  and  tangent to $\cS$;  
\item[(c)]  for any $y \in M_+$ the vector  $\bT_y$  is  non-zero and  time-like. 
\end{itemize}
 The vector field $\bT$   is    {\it the  type switching field} of the dressed barrier, while  the   component  $M_+ $,  on which $\bT$ is time-like,  is  called  the  {\it time-like region} for $\bT$. 
\end{definition}
The relevance of  this notion comes from the fact that it allows  to prove our   Theorem \ref{mainmain} in  two  steps. At  first   we show that that if $(\cS, \bT)$ is a dressed barrier, then  any oriented causal world-line, which  has     positive time-orientation  with  respect to the time-orientation ${\sT}$     on $M_+$ determined by $\bT|_{M_+}$,  
 may cross $\cS$   in  just one of the two possible directions. In the second step, we prove that if $\cS$ is a null  hypersurface (not necessarily a dressed barrier), then for any   $x_o\in \cS$ there exist a neighbourhood $\cU$ and a vector field $\bT$ on $\cU$,  which makes the pair $(\cS\cap \cU, \bT)$ a dressed barrier for  the Lorentzian manifold $(\cU, g|_{\cU})$.  From  the result of the  first step,  it immediately follows that any appropriately oriented causal world-line  may cross $\cS$ at $x_o$ in only  one direction, the other being forbidden. \par
  \medskip  
\subsection{The semi-permeability  of the dressed barriers}
The   semi-permeability   property of dressed barriers is a direct consequence of   the  following
\begin{lem}\label{main} Let   $(\cS, \bT)$  be a dressed barrier in a Lorentzian $n$-manifold  $(M, g)$, $n \geq 2$   with mostly minus signature.  Let  $\sT$ be the time orientation of $M_+$ determined by $\bT|_{M_+}$  and denote by  $[E^\perp]$   the orientation of   $\cS$,   given by  a   vector field $E^\perp$   on a neighbourhood $\cV \subset \cS$ of a point $x_o \in \cS$  satisfying the following pair of conditions:
  \beq \label{2.1} g_{x_o}(E^\perp_{x_o}, E^\perp_{x_o}) = 0\ ,\qquad  g_{x_o}(\bT_{x_o}, E^\perp_{x_o}) =- 1\ .\eeq
 Then  there is no {\rm time-like}  world-line  $ \cC$ in  $M$, which  is     $\sT$-time oriented in the region $M_+$ and  crosses  $\cS$ at $x_o$ in the  direction  of  $[E^\perp]$. 
\end{lem}
\begin{pf}  Aiming to a contradiction,   let us assume that there  is   a time-like world-line $\cC$  crossing $\cS $ at $x_o$ in  the direction of $[E^\perp]$ and    $\sT$-time oriented at the points of $ M_+$. Let   $\g:(a, b) \longrightarrow \cC$ be a global regular parameterisation  of $\cC$,  which is consonant with its orientation,   and  denote by $\cB = (e_0, e_1, e_2, \ldots, e_{n-1})$ a linear frame for $T_{x_o} M$ satisfying the following conditions:
\begin{itemize}[leftmargin = 18pt]
\item[(a)]  the vectors $e_2, \ldots, e_{n-1}$ span a space-like subspace $ W$ of $T_{x_o} \cS$ and   are  $g_{x_o}$-orthonormal; 
\item[(b)]   $e_1 \= \bT_{x_o}$ (and hence  it is  a generator for  the kernel of $g_{x_o}|_{T_{x_o} \cS \times T_{x_o} \cS}$); 
\item[(c)] $e_0$ is null,   $g_{x_o}$-orthogonal to $W$ and  such  that $g_{x_o}(e_0, e_1) = -1$. 
\end{itemize}
By  \eqref{2.1},  the  vector 
$$w \= E^\perp_{x_o} - e_0$$ 
is $g_{x_o}$-orthogonal to $e_1 =  \bT_{x_o}$ and hence  it is in the space $\bT_{x_o}^\perp  = \Span\{ e_1, e_2, \ldots, e_{n-1}\}$. It follows that, if we consider  a smooth vector field $\wt E^\perp: \cV \subset \cS \to TM$ on a neighbourhood $\wt \cV$ of $x_o$ which  satisfies $\wt E^\perp_{x_o} = e_0$, then  at the points of  a sufficiently small $\cV$  the  vector fields $E^\perp$ and $\wt E^\perp$ are  consonant and  giving  the same orientation $[\wt E^\perp] = [E^\perp|_{\cV}]$ for $\cS \cap \cV$. \par
\smallskip
  Let us  now denote by $s_o$ the value of the parameter of $\g$, corresponding to $x_o = \g(s_o)$, and let  $\dot \g(s_o) = v^0 e_0 + v^1 e_1 + \sum_{j = 2}^{n-1} v^j e_j$ be the   expansion of  $\dot \g(s_o)$  in terms of   the frame $\cB$. 
From the fact that   $\dot \g(s_o)$ is time-like and from the assumptions on  $\cB = (e_i)$,   we have that  
\beq \label{ecco} 0 < g(\dot \g(s_o), \dot \g(s_o)) = - 2 v^0 v^1 + \sum_{2 =1 }^{n-1} (v^j)^2 g_{x_o}(e_j, e_j) \leq - 2 v^0 v^1\ .\eeq
We claim that  the component $v^1$  is  positive. 
In order to check this, 
we first  observe   that, due to  the    $\sT$-time orientation of $\cC$,  for any $s \in (a, b)$  that correspond  to a point $y = \g(s) $  in  $M_+$,  the  tangent  vector  $\dot \g(s) \in T_y M$ is   in the same causal  half-cone containing  $\bT_y$.  On the other hand, for each  $s \in (s_o - \ve, s_o + \ve)$ in  a sufficiently small neighbourhood of $s_o$, we may  consider an $n$-tuple 
$\wt \cB_y = (f_0(y), f_1(y), f_2(y),  \ldots, f_{n-1}(y))$ of vectors in $T_{y = \g(s)} M$  constructed  as follows: 
\begin{itemize}[leftmargin = 18pt]
\item[--] the vectors $f_0(y), f_2(y), \ldots, f_{n-1}(y)$ are obtained by parallel transport  along the parameterised curve  $\g$ from   the vectors  $e_0, e_2,  \ldots, e_{n-1} \in T_{x_o} M$; 
\item[--] $f_1(y) \=  \bT_y$. 
\end{itemize}
Since  parallel transport preserves  scalar products, we have that
\begin{itemize}[leftmargin = 18pt]
\item[--]  $g_y(f_0(y), f_0(y)) = 0$, 
\item[--] $g_y\big(f_i(y), f_j(y)\big) = -\d_{ij}$ for any  $2 \leq i, j \leq n-1$; 
\item[--]  $g_y\big(f_0(y), f_j(y)\big) = 0$ for any $2 \leq j \leq n-1$. 
\end{itemize}
Moreover,  when  $s$ tends  to $s_o$, each vector of the  $n$-tuple  $\wt \cB_{y = \g(s)}$ tends to the corresponding element of  $\cB = (e_0, e_1, \ldots, e_{n-1})$. Thus,    if   $ \ve > 0$  is  sufficiently small,  each   $n$-tuple   $\wt \cB_{y = \g(s)}$, $s \in (s_o - \ve, s_o + \ve)$,  is  linearly independent and  is therefore  a linear frame for $T_y M$.  Due to this, a vector $w$ of a tangent space $T_{y} M$ at some  
$y = \g(s) \in M_+ \cap \g((s_o - \ve,s_o + \ve))$,   
is  time-like   if and only if it has  the form 
$w = w^0 f_0(y) +  w^1  \bT_y  + w^{2} f_{2}(y) + \ldots + w^{n-1} f_{n-1}(y)$
 with components $w^i$ satisfying the inequality
$$(w^1)^2 g(\bT_y, \bT_y) + 2 w^1 \left(  g_y(\bT_y, f_0(y))  w^0 + \sum_{k = 2}^{n-1}  g_y(\bT_y, f_k(y))  w^k\right) -  \sum_{j = 2}^{n-1} (w^j)^2> 0\ .$$
Since $ g(\bT_y, \bT_y)> 0$ for any $y \in M_+$, this  is equivalent to  
\begin{multline*}\left(g(\bT_y, \bT_y) w^1 + \left(  g_y(\bT_y, f_0(y))  w^0 + \sum_{k = 2}^{n-1} g_y(\bT_y, f_k(y)) w^k \right)\right)^2  >  \\
> \left(  g_y(\bT_y, f_0(y))  w^0 + \sum_{k = 2}^{n-1} g_y(\bT_y, f_k(y)) w^k \right)^2 + g(\bT_y, \bT_y)   \sum_{j = 2}^{n-1} (w^j)^2\ .\end{multline*}
It follows that  the two connected components of the set of non-zero time-like vectors in such tangent spaces are characterised by  either one of the following  conditions: 
\beq\begin{split}\label{23}
& g(\bT_y, \bT_y) w^1   > -  \left(  g_y(\bT_y, f_0(y))  w^0 + \sum_{k = 2}^{n-1} g_y(\bT_y, f_k(y)) w^k \right) + \\
& \hskip 2.5cm + \sqrt{  \left(  g_y(\bT_y, f_0(y))  w^0 + \sum_{k = 2}^{n-1} g_y(\bT_y, f_k(y)) w^k \right)^2+ g(\bT_y, \bT_y)   \sum_{j = 2}^{n-1} (w^j)^2}\ ,\\
&\hskip 6 cm \text{or}
\end{split}
\eeq
\beq\begin{split}\label{24}
& g(\bT_y, \bT_y) w^1   <  - \left(  g_y(\bT_y, f_0(y))  w^0 + \sum_{k = 2}^{n-1} g_y(\bT_y, f_k(y)) w^k \right) - \\
& \hskip 2.5 cm - \sqrt{  \left(  g_y(\bT_y, f_0(y))  w^0 + \sum_{k = 2}^{n-1} g_y(\bT_y, f_k(y)) w^k \right)^2+ g(\bT_y, \bT_y)   \sum_{j = 2}^{n-1} (w^j)^2}\ .
\end{split}
\eeq
Since  at the  points $y \in M_+$  the right hand side of \eqref{23}  is non-negative, while the right hand side of \eqref{24}  is non-positive,    the   half-cones  of the time-like vectors are distinguished   by    the condition  $w^1  > 0$ or  the condition $ w^1  < 0$.  This implies that,  for any  $s \neq s_o$ in a  sufficiently small neighbourhood of  $s_o$ and   with $y = \g(s) $ in  $M_+$, 
 the condition  that the vector  $\dot \g(s)$  is in the same causal  half-cone of $\bT_{y = \g(s)}$ is equivalent  to demanding  that its component $w^1$ in the direction of the vector  $f_0(y) = \bT_y$ is strictly positive. \par
We are now ready to conclude the proof that  the component $v^1$  of $\dot \g(s_o)$ is positive. Indeed, if $v^1 < 0$, by continuity, the same should hold for  the  component $w^1$ of any tangent    vector $\dot \g(s)$, with  $s\neq s_o$ sufficiently close to  $s_o$. But this would  contradict the above remarks and the  hypothesis  on the  time orientation of $\cC$.  Combining this with the fact that  $v^1 \neq 0$ by  \eqref{ecco},  we get that    $v^1 >0$.
\par
We may now conclude the proof. Indeed, since $v^1 > 0$,   \eqref{ecco} implies that   $v^0 < 0$ and thus  that  $\cC$  crosses $\cS$  in the direction  $[- \wt E^\perp]$,  contradicting  the hypothesis  that $\cC$ crosses $\cS$ at $x_o$  in the direction of $[\wt E^\perp] =  [E^\perp]$. 
\end{pf}
This result has the following corollary, which concludes the first step. 
\begin{cor} \label{thecor} Under the  hypotheses of Lemma \ref{main}, there is no {\rm causal} world-line $\cC \subset \cU$, which  is     $\sT$-time oriented and  crosses  $\cS$ at  $x_o$ in the  direction  of  $[E^\perp]$.
\end{cor}
\begin{pf}  We only need to show  that there is no causal world-line $\cC$ which is  null   at  $x_o$,    $\sT$-time oriented and crossing  $\cS$  in  the direction of $[E^\perp]$.   Assume on the contrary that such a curve does exist and pick a regular parameterisation  $\g: (s_o - \ve, s_o + \ve) \to \cI \subset  \cC$  for a neighbourhood $\cI$  of $x_o = \g(s_o)$.  Denote $v_o \= \dot \g(s_o) \neq 0$.  Since $v_o$ is null, there exists    a sequence of  time-like vectors $v_k \in T_{x_o} M$ with $\lim_{k \to \infty} v_k = v_o$. 
Using the same frame $\cB = (e_0, \ldots, e_{n-1})$ for $T_{x_o} M$   of  the proof of Lemma  \ref{main}, we get that  $v_o$ and the $v_k$'s  uniquely decompose as $v_o   = \lambda_o E^\perp_{x_o} + w$ and  $v_k = \lambda_k E^\perp_{k_o} + w_k$, respectively,   for some  $w, w_k \in T_{x_o} \cS$. Since  $\lambda_o > 0$ by hypothesis,    we also  have that   $\lambda_k > 0$   for any $k$ sufficiently large.  Working in  coordinates,  one can  construct  a sequence of  smooth parameterised curves $\g_k: (s_o- \ve, s_o + \ve) \to \cU$, which  converge uniformly on closed intervals to  the map $\g: (s_o - \ve, s_o + \ve) \to \cU$ and  such that   $\g_k(s_o) = x_o$,   $\dot \g_k(s_o) = v_k$. By construction,   for  $k$ sufficiently large, there are neighbourhoods $(s_o - \ve_k, s_o + \ve_k)$ of $s_o$  where the  restrictions $\g_k|_{(s_o - \ve_k, s_o + \ve_k)}$  are regular parameterisations of time-like world-lines and with the property that such parameterisations are consonant with the  $\sT$-time orientation of   $\cU_+$.  Thus,  all such  curves are time-like world-lines, which are    $\sT$-time oriented and   crossing $\cS$  in the  direction  $[E^\perp]$. This cannot be by  Lemma \ref{main}.\end{pf}
\par
\medskip
\subsection{The semi-permeability property of  time-oriented null hypersurfaces}
The  second step  of our proof  basically  consists  of the following 
\begin{lem} \label{34} Let $(M, g)$ be  a Lorentzian $n$-manifold  with mostly negative signature and  $\cS \subset M$  a  null hypersurface, which is time-oriented by  a nowhere vanishing null vector field $E \in T\cS$.   For any $x_o \in \cS$,  there exist a neighbourhood $\cU \subset M$  of $x_o$ and a vector field $\bT$ on  $\cU$,  such that $\bT|_{\cU \cap \cS} = E|_{\cU}$  and  $(\cS \cap \cU, \bT)$ is  a dressed barrier for the Lorentzian manifold $(\cU, g|_\cU)$.
\end{lem} 
\begin{pf}  Consider a  neighbourhood $\cU$ of $x_o$,  on which there is  a nowhere vanishing null vector field $\wt E$  which coincides with  $ E|_{\cU \cap \cS}$ at  the points of $\cU \cap \cS$.  We may also assume that on $\cU$ there is  a null vector field $\wt E^\perp$   satisfying the condition  $g(\wt E, \wt E^\perp) \equiv  - 1$ and  that there is  a system of coordinates $(x^1, \ldots, x^{n-2}, u, v)$ on $\cU$,   such that  $\cS \cap \cU = \{v = 0\}$. 
Let 
$$\cU_+ =  \{ x \in \cU\ :\ v > 0\}\ ,\qquad \cU_- =  \{ x \in \cU\  :\ v < 0\}$$ 
and denote by $\bT$ the vector field on $\cU$ defined by 
\beq  \bT = \wt E - v  \wt E^\perp\ . \eeq
By construction, for any $x \in \cS \cap \cU$,  the vector $\wt E_x = E_x$ is tangent to $\cS$ and for any $y \in ( \cS \cap \cU) \cup \cU_+$ we have that 
 \begin{multline*}g_y(\bT, \bT) =- 2  v(y) g_y(\wt E, \wt E^\perp) = 2 v(y) \geq 0\ \\
 \text{with}\  g_y(\bT, \bT) = 0 \qquad \text{if and only if}\ y \in \cS \cap \cU\ .\end{multline*} 
It follows that  $(\cS \cap \cU, \bT)$ satisfies all conditions to be  a dressed barrier for $(\cU, g|_\cU)$. 
\end{pf}
 We are now able to  prove  our main result. \par
  \begin{theo} \label{mainmain}   Let $(M, g)$ be  a Lorentzian $n$-manifold  with mostly negative signature and  $\cS \subset M$   an orientable null hypersurface,    equipped with the time-orientation ${\sT}^{(\cS)}$ determined by a 
  nowhere vanishing  null vector field  $E \in T \cS$. Denote   by   $E^\perp \in TM|_{\cS} \setminus T \cS$   a null vector field such that  
  \beq \label{thecond} g_x(E^\perp, E) =  -1\qquad \text{for any}\ x \in \cS\ .\eeq 
  Given  $x_o \in \cS$ and  a neighbourhood $\cU \subset M$ of $x_o$, equipped with  a vector field $\bT$  as in Lemma \ref{34} and  with the  time-orientation $\cT$ on $\cU_+$ determined by $\bT$,    
  no oriented causal world-line $\cC \subset M$ 
  crosses  $\cS$ at $x_o$  in the direction of $[E^\perp]$ (resp. $[-E^\perp]$) if its orientation is compatible with the  $\cT$-time-orientation (resp.  $(-\cT)$-time-orientation) in    $\cU_+$. 
 \end{theo}
 \begin{pf} If the  orientation is compatible with  the  $\cT$-time-orientation,  the arc  $\cC' = \cC \cap \cU$    is  a $\cT$-time-oriented causal world-line for the Lorentzian manifold $(\cU, g|_\cU)$ and  cannot cross  $\cS \cap \cU$ in the direction of $[E^\perp]$ by  Corollary \ref{thecor}.  An analogous  conclusion occurs when the  orientation is compatible with  the  $(-\cT)$-time-orientation.
 \end{pf}\par
\medskip
\section{A few (classical and not) examples of black holes and barriers}
\label{examples}
As we defined in the Introduction,   a  null hypersurface  of  a Lorentzian $n$-manifold $(M, g)$ is called   {\it (semi-permeable) barrier}   if (a)  it is equipped with a time-orientation $\sT^{(\cS)}$ determined by a no-where vanishing null vector 
field $E$ in $T\cS$, and (b)  separates $M$ into two connected components, i.e.   $M \setminus \cS = M_- \cup M_+$  where $M_+$ and $M_-$ are disjoint   connected regions.  \par
%
\smallskip
By Theorem \ref{mainmain},   {\it if the signature of $g$ is mostly minus and if $E^\perp$ is a null vector field at the points of a barrier $\cS$, which  takes values in  $TM|_{\cS} \setminus T\cS$  and  satisfies  \eqref{thecond}, then 
no appropriately oriented causal world-line can cross the  barrier $\cS$ in the direction of $[E^\perp]$}.  On the other hand, {\it in  case   the signature is mostly plus,  by just reversing the sign of the metric, one immediately gets  that  the forbidden direction   is $[- E^\perp]$} (instead of $[E^\perp]$). \par
\smallskip 
Note that these two  rules give an  immediate way to determine whether a smoothly bounded region $\O$  in the  complement $M \setminus \O_o$ of a time-oriented region $\O_o$  is  {\it black}  or {\it white}  for $\O_o$ according to the definitions given in the Introduction.
\par
\medskip
 As we mentioned above, the semi-permeability property of the barriers is the well known crucial property of  Penrose, Hawking and Ellis' event horizons (see  Appendix \ref{PHE}).  Indeed,  our main result
 provides an extremely simple method to determine the event horizons for all  known examples of static and  Vaidya-type black holes.  In this section we illustrate it   in the cases of   Kerr-Newman black holes and a certain kind of Myers-Perry black holes.  One can directly check  that  the method  works equally well for any other known models  of black holes.  Finally, we  will use the same method to determine very quickly the barriers for the large two new classes of Einstein metric recently determined in \cite{GGSS} and which naturally generalise the event horizons of  the classical Kerr metrics of rotating black holes. \par
 \smallskip
   Before  introducing  these various examples, we just need to  recall that  one of the  simplest way to check  whether a given  hypersurfaces  $\cS$  is time-like (resp. space-like, null) is to verify  whether,  for any $x \in \cS$,  there exists  a locally defined  real  function $F: \cU \subset M \longrightarrow \bR$ on a neighbourhood $\cU$ of $x$ satisfying the following two conditions
\begin{itemize}
\item   $\cS \cap \cU = \{F = R\}$  for some constant $R$ and 
\item the $1$-form  $dF$ is  time-like (resp. space-like, null) at the points of $\cS$ i.e. 
\beq \label{test} g^{-1}(dF, dF)|_{\{F= R\}} > 0\qquad \bigg(\ \text{resp.}\  < 0\ ,\ \ = 0\ \bigg)\ .\eeq
\end{itemize}
\par
\subsection{The  inner and outer barriers of  Kerr-Newman black holes}\label{sect41}
The Kerr-Newman metrics (\cite{NJ, NCCEPT, Ca1, DKS}) can be considered as the Lorentzian metrics $g_{KN}$  on the  manifold $M = (\bR^3 \setminus \{0\})\times \bR$,   which take the following form in   cylindrical  {\it Kerr coordinates}   $((\xi, \psi, \r);  v)$  (see e.g. \cite[eq. (7.4)]{DKS}) (\footnote{ Note that \eqref{KerrNewmann} is  not exactly  the  metric  appearing in   \cite{DKS}, but it is scaled by  the factor $- \frac{1}{2}$. In this way  the signature changes from mostly plus to mostly minus and  the metric is normalised in a way that  allows an easier comparison with the metrics in \cite{GGSS}, which are discussed later in this section.}).
\begin{multline}  \label{KerrNewmann}
 g_{KN} \=  - \frac{1}{2}(\r^2 + a^2 \cos^2\! \xi) \left( d \xi^2 + \sin^2\! \xi d \psi^2\right) - \\
  -   \left(dv - a \sin^2\! \xi d \psi\right)\vee \left(d \r   - a \sin^2\! \xi   d \psi\right) +  \frac{1}{2}  \left(1 - \frac{2 m \r - e^2}{\r^2 + a^2 \cos^2\! \xi} \right) \big(d v - a \sin^2 \!\xi d\psi\big)^2 
\ , 
\end{multline}
where $m$, $a$ and $e$ are three constants, which are physically interpreted as mass, angular momentum per mass unit and electric charge, respectively,  of the black hole. 
We discuss just  the case  in which  $m$, $a$ and $e$  satisfy the inequality 
$a^2 + e^2 < m^2$. 
Under this assumption, let us  look for a  null hypersurface of the form  $\cS_R  = \{  \rho = R \}$ for some real constant $R$.  Such a hypersurface (if it exists)  is clearly diffeomorphic to $S^2 \times \bR$ (and thus orientable and time-orientable) and separates $M =   (\bR^3 \setminus \{0\})\times \bR$ into two connected components. In particular {\it any such  hypersurface, equipped with a  time-orientation,   is  a barrier}.\par
\smallskip
  A simple computation shows that  the   real function $F(\xi, \psi, \r, v) \= \r $  is such that  for any real  constant $R$ 
 $$g_{KN}^{-1}(dF, dF)\big|_{ \{F = R\}} = 2\frac{- a^2 - R^2+ 2m  R - e^2}{R^2 + a^2 \cos^2 \xi}\ . $$
 Thus, the hypersurface  $\cS_R = \{ F = R\}$  is null if and only if $R$ makes the numerator equal to  zero, i.e.  if and only if  $R = R_\pm$ with 
 $$\text{either}\qquad R_+ = m + \sqrt{m^2 - a^2 - e^2}\ \qquad \text{or} \qquad R_- = m -  \sqrt{m^2 - a^2 - e^2}\ .$$
 By the above remarks $\cS_{R_+}$ and $\cS_{R_-}$ are  barriers with  null distributions  generated by the vector fields
 \begin{multline} E_{(+)}\=- \frac{\p}{\p \psi}+ \frac{2}{a} (m + \sqrt{m^2 - a^2 - Q^2})\frac{\p}{\p v}\bigg|_{\cS_{R_+}}\ ,\\ E_{(-)} \= -  \frac{\p}{\p \psi}+\frac{2}{a} (m - \sqrt{m^2 - a^2 - Q^2})\frac{\p}{\p v} \bigg|_{\cS_{R_-}}\ ,\end{multline}
 respectively. 
 Since  $\frac{\p}{\p \r}$ is   transversal to $\cS_{R_+}$ and   such that 
 $g\left(\frac{\p}{\p \r}, E_{(+)}\right) < 0$,  it follows that  $\frac{\p}{\p \r}\big|_{\cS_{(+)}}$  and the vector field $E^\perp_{(+)} \in TM|_{\cS_{(+)}} \setminus T\cS_{(+)}$ defined in \eqref{thecond} 
point towards the same side of $\cS_{(+)}$. Hence if we time-orient $\cS_{(+)}$ using the vector field  $E_{(+)}$, by Theorem \ref{mainmain} we get that $\cS_{(+)}$  is a barrier for $(M, g_{KN})$ with the property that  no signal can pass from the region $\O_1 \=  \{\r < R_+\}$ into the region $\O_o \= \{\r > R_+\}$. In fact, {\it $\cS_{(+)}$ is nothing but the well known outer event horizon of the Kerr-Newman black hole}.\par
A similar discussion shows that $\cS_{(-)}$, equipped with either one of its two possible time-orientations, is a barrier. In fact it is what is usually called {\it inner horizon}.
%
\subsection{Barriers and  horizons of   Myers and Perry's  n-dimensional black holes}
One of the simplest version of  a Myers and Perry's spinning uncharged $n$-dimensional   black holes  are given by the  Lorentzian metrics $ g_{MP} $ on the manifold   $M = S^{n-4} \times (\bR^3 \setminus \{0\}) \times \bR$, which in  coordinates $((\theta_1, \ldots, \theta_{n-4}), (\xi, \psi, \r), v)$, with $\theta_i$ coordinates for $S^{n-4}$ and  $(\xi, \psi, \r)$ spherical coordinates for $\bR^3 \setminus \{0\} $,  take the form  (see \cite[eq. (3.1)]{MP})  (\footnote{As   for  \eqref{KerrNewmann},   this is   the   expression given  in \cite{MP}  scaled by the factor   $- \frac{1}{2}$.}).
\begin{multline} g_{MP}  =- \frac{1}{2} (\r^2 + a^2 \cos^2\!\xi) (d xi^2 + \sin^2 \!\xi \,d \psi^2) + \r^2 \cos^2\! \xi \, g_{S^{n-4}} - \\
-  ( dv - a \sin^2 \!\xi \, d \psi )\vee (d\r - a \sin^2\! \xi  \, d \psi) 
+  \frac{1}{2} \left( 1 - \frac{m}{\r^{n-5}(\r^2 + a^2 \cos^2\! \xi)}\right) (dv - a \sin^2\! \xi d \psi)^2 
\ , \end{multline} 
where  $m$ and $a$ are constants, physically interpreted as mass and angular momentum per mass unit  of the black hole, and $g_{S^{n-4}}$ is the standard round metric of $S^{n-4}$.  The time-oriented null hypersurfaces  of the form  $\cS_R  = \{  \rho = R \}$ for some real constant $R$   are barriers by the same observations  we made for the Kerr-Newman metric. These hypersurfaces are those for which  the function $F(\theta_i, \xi, \psi, \r, v) {\=} \r$ satisfies the condition  
$ g^{-1}_{MP}(dF, dF)\big|_{\{F {=} R\}} {\equiv} 0$.  A direct check shows that this condition  has  the form 
\begin{multline} \label{equa} 0 = g_{MP}^{-1}(dF, dF)\big|_{ \{F = 0\}} = G(\theta_i, \xi, \psi, R) (R^{n-5}(R^2 + a^2) - m) \\ \text{where}\ G(\theta_i, \xi, \psi, R)\ \text{is an appropriate nowhere vanishing function  on}\ M\ . \end{multline}
Since  $f(R)\=R^{n-5}(R^2 + a^2) - m$ is monotone  increasing on $[0, + \infty)$ and  with $\lim_{R \to \infty} f(R) = + \infty$, there is
either none or exactly one positive solution to  the equation \eqref{equa}, depending on whether $\lim_{R \to 0^+} f(R)$ is non-negative or negative, respectively.  If $n \geq 6$, this limit is surely  negative (as it has been observed  in \cite{MP}), while if  $n =5$, the limit  is negative  if and only if $a^2 < m$.  In these cases, the unique solution $R_o$ to \eqref{equa}  determines a barrier $\cS_{R_o}$. Indeed it is the event horizon of  a  Myers and Perry's black hole (or of a Myers and Perry's  white hole, depending on which  time-orientation is considered).\par
\smallskip
Similar  discussions  allow to determine the   event horizons of  Myers and Perry's black  holes  for  other  metrics considered    in \cite[\S\ 3.1]{MP}.
\par
\medskip
\subsection{Barriers  in Ricci flat  manifolds with Kerr type optical structures} Let $M$ be a $4$-dimensional manifold of the form 
$M = \cU \times \bR^2\subset \bC \times \bR^2$,  with  $\cU \subset \bC$  open and simply connected, equipped with coordinates $(x, y, v,\r)$ 
among which $x, y$ are the real and imaginary parts of the standard complex coordinate $z = x + \I y$ of $\cU \subset \bC$ and $(v, \r)$ are the standard coordinates
of $\bR^2$.  Denote also by $\k$ an integer which can be either $+1$ or $-1$ and let $m\geq 0$ be a non-negative  real constant. 
 In \cite{GGSS} it was proved that, for any nowhere vanishing   $\f: \cU \rightarrow \bR$  with    $\k   \f < 0$ and solution to    the elliptic equation
\beq \label{firsteq**}   \Delta \f  + \k \frac{8 \f}{\left(1 + \k (x^2 + y^2)\right )^2} =0\ ,\eeq
the   Lorentzian metric on $M$  
  \begin{multline} \label{comp2}
 g^{(\k, \f)}\=   -  2 \k  \frac{\r^2 + \f^2  }{(1+\k (x^2 + y^2))^2} (dx^2 + dy^2) +\\
 + \big(d v - \frac{\p \f}{\p y} dx +  \frac{\p \f}{\p x} dy\big) \vee \bigg( dv + d\r + \frac{1}{2}\left( - 1 + \frac{ 2m \, \r }{\r^2 + \f^2} \right)\big(d v - \frac{\p \f}{\p y} dx +  \frac{\p \f}{\p x} dy\big) \bigg)\ ,
  \end{multline}
is  Ricci flat.\par
\smallskip
  The  Einstein metrics of this kind with  $\k = +1$   constitute a very  large class, which  properly  include the  metrics of the Kerr black holes. Those  with $\k = -1$ give an equally large class and  can be taken as  their ``hyperbolic'' analogs. Both classes have been characterised in \cite{GGSS} as the only Ricci flat metrics which are compatible with a special type of   optical structures on $M$, called {\it of Kerr type}, and satisfy a natural set of conditions  that encode some of  the most important features of Kerr metrics (see \cite{GGSS} for details).  \par
  \smallskip
 Explicit solutions to \eqref{firsteq**} -- and, consequently,  explicit expressions for the corresponding metrics   $g^{(\k, \f)}$ -- can be  determined as follows.  It was remarked in \cite{GGSS} that  the solutions to   \eqref{firsteq**}   on  a  disk $\overline{\bD(0,\gr_o)}\= \{\sqrt{x^2 + y^2}\leq r_o\} $, $0 < r_o < 1$,   are real analytic up to the boundary and in one-to-one correspondence with  the Fourier series  of real analytic  boundary data on $\p \bD(0,r_o)$. 
     In fact,  using  polar coordinates $(r , \psi)$   for  the $(x, y)$-plane  (so that $(x = r \cos \psi, y =  r \sin \psi)$) and  considering the Fourier expansion $f(\psi) = a_0 + \sum^\infty_{n = 1}( a_n  \cos(n \psi) +  b_n \sin(n \psi))$ of a non-negative real analytic  function  $f(\psi)$ on $\partial \bD(0,  r_o) $, $r_o < 1$,   the corresponding solution $\f$  to \eqref{firsteq**}  with $\k = + 1$ (resp. $\k = -1$)  is  the sum of the  series 
     \beq\label{explicita}
     \begin{split}
     & \f(r, \psi) = -  \left( a_0  \frac{\f_0(r)}{\f_0(r_o)} + \sum^\infty_{n = 1}\frac{\f_n(r)}{\f_n(r_o)}\big(  a_n \cos(n \psi) +   b_n \sin(n \psi)\big)\right)\ ,\\
     & \text{with}\ \ \ \f_n(r) \=   \left(1 - \frac{2 r^2}{ (1 + r^2)}\frac{1}{n+1} \right)  r^n \ \ 
     \bigg( \text{resp. }\ \ \  \f_n(r) \=\left(\frac{1+r^2}{1 - r^2} + n\right) r^n\ \ \bigg)\ .
     \end{split}
     \eeq
Up to isometries, the classical Kerr metrics   are  those  of  the family with  $\k = + 1$  and for which the function   $\f$ is   as in  \eqref{explicita} with  $a_n = b_n = 0$ for any $n \geq 2$ (\cite[\S 6.2.1]{GGSS}).\par
\medskip
Inspired by  how the event horizons of Kerr-Newman metrics can be determined, one can look for barriers of the form $\cS_R = \{ F = 0\}$ for some real $F$ of the form 
$$F(x, y, \r, v) \= \r - R(x, y, v)\qquad \text{for a smooth function of three variables} \ R(x, y, v)\ .$$
 These hypersurfaces are barriers  if and only if  $R(x, y,v)$ satisfies the first order differential constraint
\begin{multline} \label{constr}
g^{(\k, \f)-1}\bigg(\frac{\p R}{\p x} dx + \frac{\p R}{\p y} dy  + \frac{\p R}{\p v} dv,\frac{\p R}{\p x} dx + \frac{\p R}{\p y} dy  + \frac{\p R}{\p v} dv\bigg)\bigg |_{\r= R(x, y, v)}  =\\
=  g^{(\k, \f)-1}(d\r, d\r ) \big |_{\r= R(x, y, v) } \ .\end{multline}
This is a real analytic differential equation,  which is  quadratic in the partial derivatives $\frac{\p R}{\p x}$, $\frac{\p R}{\p y}$  and $\frac{\p R}{\p v}$ and admits at least two distinct solutions if $\k = +1$ and $\f(x, y)$ corresponds to  a Kerr metric.  It is therefore reasonable to expect that whenever $\f(x, y)$ is sufficiently close (in an appropriate norm) to  a function for  a Kerr metric, then \eqref{constr}  still admits two distinct solutions, possibly diverging at some points.  \par
\smallskip
On the other hand, we recall that  the definition of  barriers in a Lorentzian manifold is completely coordinate-free. This yields that  the class of barriers is  preserved by any  isometry of the space-time that contains them.  Thus  studies on  the (local) existence and on the geometric properties of the solutions to \eqref{constr} might determine useful information on the  space-times $(M, g^{(\k, \f)})$ and how  the  event horizons of classical  Kerr black holes might be deformed.  Investigations of this kind are left to  future work. 
\par
\medskip
\appendix
\section{Penrose, Hawking and Ellis' definitions  of black holes and  event horizons}\label{PHE}
 Here is a very short  presentation of  the   notions  of  {\it black hole} and  {\it  event horizon}, as  it can be found in   classical  papers and textbooks as  e.g. \cite{Pe, HE, Wl, Ca2, Ca3}.  At the end, we  also  give a  direct proof of the fact that any smooth event horizon is necessarily a null hypersurface.  \par
\smallskip
Let  $(M, g)$ be a Lorentzian $n$-manifold admitting  a conformal embedding 
$$\imath: (M, g) \longrightarrow (\cM, \gg)$$
 into a larger Lorentzian manifold $ (\cM, \gg)$   with a relatively compact  image $\imath(M)$ (\footnote{The existence of such a conformal embedding  with compact closure $\overline{\imath(M)}$  guarantees  that  each oriented causal  world-line  of $(M, g)$ is mapped by the embedding $\imath$ into an oriented causal world-line $\cC$ of   $(\cM,  \gg)$ with a well-defined  upper  limit end-point $p^{(\cC, \infty)}$ in $\cM$. Such upper limit $p^{(\cC, \infty)}$  can be taken as a mathematical object that properly encodes     the  intuitive notion   of  the ``infinite future point''  of the causal world-line  $\cC$.}). Assume that $(M, g)$   satisfies    conditions  that guarantee   that it is    {\it asymptotically flat at null and spatial infinity}.  For an explicit detailed  lists of   conditions of this kind  see e.g.    \cite[p. 276]{Wl}  and \cite[\S II.2]{Ca3}.
Let us now fix  a  time-oriented region $\wt \O_o$  of $M$ and  denote by  $\cI^+(\wt \O_o)$  the  set of the points in  $\overline{\imath(M)} $, which are   final  end-points of the (traces of the) time-oriented  inextensible null geodesic of $(M, g)$  starting  from points of $\wt \O_o$. Finally, let us  denote  by    $\O_o \= J^-(\cJ^+(M_o))$   the {\it causal past} of  $\cI^+(\wt \O_o)$, i.e. the subset  of points of $M$ that  can be joined to  points in $\cI^+(\wt \O_o)$ through polygonals of positively time-oriented causal geodesics. \par
\smallskip
 Following  Carter's terminology  \cite{Ca2},  we  call  such a region $\O_o$ the  {\it domain of outer communication determined by  $\wt \O_o$}. We are now ready to state  the following (very minor) variant of  {\it Penrose, Hawking and Ellis' definitions for black holes and event horizons} (see e.g. \cite[p.300]{Wl}). \par 
\begin{definition} \label{PHEdef}
The {\it black holes  for the observers in  $\wt \O_o$}  are the (interiors of the) connected components of the complement   $\O_o^c \= M \setminus \O_o$ of the domain of outer communication $\O_o$ determined by $\wt \O_o$.    An {\it event horizon for $\wt \O_o$}  is the  boundary  in $M$   of a  black hole.
\end{definition}
 By a lemma  of   Penrose on ``achronal boundaries'' (\cite[Lemma I at p.\ 214]{Pe} or \cite[Lemma\ 6.3.2]{HE}), any event horizon is generated by segments of null geodesics. This immediately implies that  {\it if an event horizon is a (smooth) hypersurface, then such hypersurface is null}.  \par
 \smallskip
Such a crucial    property    of   smooth event horizons  admits   the following  alternative,  quite elementary proof. First of all, it is important to observe that,  as an immediate consequence of   Definition  \ref{PHEdef}, if an event horizon $\cS \subset M$  is a smooth hypersurface   which bounds a black hole $\O$,  then for any neighbourhood $\cU \subset \cS$ of a point  $x_o \in \cS$,  no physical (causal) signal  might cross $\cU$ passing from the black hole to the domain of outer communication. This means that such $\cU \subset \cS$ has  the semi-permeability property  with respect to time-oriented causal world-lines described in Theorem \ref{mainmain}.  By  the discussion in the Introduction,  this  implies  that   $\cU$  cannot  be  time-like (Fig.3).  We claim that  $\cU\subset \cS$ cannot  be space-like either. In fact,  otherwise  a sufficiently small smooth deformation $\cU'$ of  $\cU$ would be part of  a new space-like hypersurface $\cS'$,  with the following two properties: (a) it   still has  the  semi-permeability property with respect to signals and  (b) it  bounds a  region $\O'$, which properly includes the black hole $\O$ and with no intersection with the domain of outer communication $\O_o$. This would contradict the assumption that $\O$ is a connected component  (= connected subset which is maximal with respect to inclusion) of $\O_o^c = M \setminus \O_o$.  We  conclude that $\cS$ is necessarily a null hypersurface. \par

\vskip 1.5truecm
\hbox{\parindent=0pt\parskip=0pt

\vbox{\baselineskip 9.5 pt \hsize=3.1truein
\obeylines
{\smallsmc
 Cristina Giannotti \ \&\ Andrea Spiro
Scuola di Scienze e Tecnologie
Universit\`a di Camerino
Via Madonna delle Carceri
I-62032 Camerino (Macerata)
Italy
}\medskip
{\smallit E-mail}\/: {\smalltt cristina.giannotti@unicam.it
}
{\smallit E-mail}\/: {\smalltt andrea.spiro@unicam.it
}
}
}

\medskip

 



 \begin{thebibliography}{XXX}
 
 \bibitem{AN} T.\ Adamo and E.\ T.\ Newman, {\it The Kerr-Newman metric: A Review},  arXiv:1410.6626 (2016).
 
 \bibitem{ABHP}  S.\ $\overset{\circ}{\textrm{A}}$minneborg, I.\ Bengtsson, S.\  Holst and P. Peld\'an,
 {\it Making anti-de {S}itter black holes},
Classical Quantum Gravity  {\bf 13},
(1996), 2707--2714.
 
 \bibitem{ANN}
 P.\ Anninos, D.\ Bernstein, S.\ Brandt, J.\  Libson, J.\  Masso, E.\ Seidel,  L.\ Srnarr, W.-M. Suen and P. Walker,
 {\it Dynamics of apparent and event horizons}, 
 Phys. Rev. Lett. {\bf 74} (1995),  630 – 633.
 
 
\bibitem{BHTZ} M.\ Ba\~nados, M.\ Henneaux, C.\ Teitelboim and J.\  Zanelli,
 {\it Geometry of the {$2+1$} black hole},
Phys. Rev. D (3)  {\bf 48} (1993), 1506--1525.

\bibitem{BS} T.\ W.\ Baumgarte. and S.\ L.\ Shapiro,  {\it Numerical relativity and compact binaries},
{Phys. Rep.}  {\bf 376} (2003), 41--131.

\bibitem{BM} I. Booth and J. Martin, {\it Proximity of Black Hole Horizons: Lessons from Vaidya spacetime}, 
Phys. Rev. D 82 (2010),  124046. 

\bibitem{BL} R. H. Boyer and W. L. Lindquist, {\it Maximal Analytic Extension of the Kerr metric}, 
J. Math. Phys. {\bf 8} (1967), 265--281.

\bibitem{BLP} D.\ R.\ Brill, J. Louko and P. Peld\'an, {\it Thermodynamics of (3+1)-dimensional black holes with toroidal or higher genus horizons}, Phys. Rev. D {\bf 56} (1997), 3600.



 \bibitem{Ca0}, B.\ Carter, {\it Hamiltom-Jacobi andSchr\"odinger Separable Solutions of Einstein's Equations}, 
 Commun. math. Phys. {\bf 10} (1968), 280--310.
 
 \bibitem{Ca1}
 B.\ Carter,  {\it Global structure of the Kerr Family of Gravitational Fields}, 
 Phys. Rev. {\bf 174} (1968), 1559--1571.
 
 \bibitem{Ca2}
 B.\ Carter,  Republication of: {\it {B}lack hole equilibrium states. {P}art {I}.
              {A}nalytic and geometric properties of the {K}err solutions}
Gen. Relativity Gravitation {\bf 41},
(2009), 2873--2938. 

\bibitem{Ca3}
B.\ Carter, Republication of: {\it {B}lack hole equilibrium states. {P}art {II}.
              {G}eneral theory of stationary black hole states},
Gen. Relativity Gravitation  {\bf42}
(2010). 653--744.
 
 \bibitem{Ch} P.\ T.\ Chru\'sciel,
 {\it Black holes---an introduction}, in 
``100 years of relativity'', pp. 93--123,
{\it World Sci. Publ., Hackensack, NJ},
2005. 


\bibitem{CN} A.\ Coudray and J.-P. Nicolas, 
{\it Geomtry of Vaidya spacetimes}, 
Gen. Relativity Gravitation  {\bf 53} (2021), Paper No. 73, 23.

\bibitem{DKS} G. Debney, R. P. Kerr and A. Schild, 
{\it Solutions of the Einstein and Einstein-Maxwell Equations}, 
J. Math. Phys. {\bf 10} (1969), 1842--1854.


\bibitem{Do} V.\ I.\  Dokuchaev,  {\it 
Is there life inside black holes?}, 
Class. Quantum Grav. {\bf 28} (2011), 23501. 



\bibitem{DB} 
K. L. Duggal and A. Bejancu,
Lightlike submanifolds of semi-{R}iemannian manifolds and
              applications,
 {\it Kluwer Academic Publishers Group, Dordrecht}, 1996. 
 


\bibitem{GGSS} 
M. Ganji,
C. Giannotti, 
G. Schmalz and A. Spiro,
{\it Einstein manifolds  with  optical geometries of Kerr type},
Ann. Physics,
{\bf 474} (2025),  Paper No. 169908.

\bibitem{GT} M. G\"urses and B. Tekin,
{\it Kerr-Vaidya type radiating black holes in semi-classical gravity with conformal anomaly}, 
Phys. Rev. D {\bf 109} (2024),  024001


\bibitem{HE} S.\ W.\ Hawking and G.\ F.\ R.\ Ellis,
 {The large scale structure of space-time},
 {\it Cambridge University Press, Cambridge}, 2023.
 
 
\bibitem{JKC} J. Jezierski,  J.  Kijowski and E. Czuchry,
 {\it Geometry of null-like surfaces in general relativity and its
              application to dynamics of gravitating matter},
    Rep. Math. Phys. {\bf 46},
(2000), 399--418. 


\bibitem{KMV}
D.\ Klemm, V.\  Moretti and L.\  Vanzo,
{\it Rotating topological black holes},
Phys. Rev. D (3)  {\bf 57}
(1998), 6127--6137. 






\bibitem{MP} R. C. Myers and M. J. Perry, {\it Black Holes in Higher Dimensional Space-Times}, Ann. Physics  {\bf 172} (1986), 304--347.

\bibitem{LMSSW}   J. Libson, J. Mass\'o, E. Seidel, W.-M. Suen and P. Walker, 
 {\it Event horizons in numerical relativity: Methods and tests}, Phys. Rev. D
 {\bf 53} (1996), 4335--4350.

\bibitem{NJ} E. T. Newman and A. I. Janis, {\it 
Note on the Kerr Spinning‐Particle Metric}, 
J. Math. Phys.  {\bf 6} (1965),  915--917.

\bibitem{NCCEPT} E. T. Newman,  E. Couch,  K. Chinnapared, A. Exton, A. Prakash, R. Torrence, 
{\it Metric of a Rotating, Charged Mass}, J. Math. Phys. {\bf 6} (1965),  918--919.

\bibitem{ON} B.  O'Neill,  Semi-{R}iemannian geometry,  {Academic Press, Inc. 
              New York},
(1983).

\bibitem{Pe} R. Penrose, {\it Structure of space-time}, 
in Battelle rencontres - 1967 lectures in mathematics and physics: Seattle, WA, USA, 16 - 31 July 1967'', 
   C.M. Dewitt and J.A. Wheeler, ed., 
   {\it Princeton U. Press, Princeton}, 1968.
   



\bibitem{St} I. Sternberg, Lectures on Differential Geometry, {\it Chelsea Publishing Co., New York}, 1983.




\bibitem{Va} L.\ Vanzo, {\it Black holes with unusual topology},
Phys. Rev. D (3)  {\bf 56}
(1997), 6475--6483.

\bibitem{Wl} R.\ M.\ Wald, General Relativity, {\it The Chicago University Press, Chicago and London}, 1984.
\bibitem{Wa} F.\ W.\ Warner, Foundations of Differentiable Manifolds and Lie groups, {\it Springer-Verlag, New York}, 1983.

\end{thebibliography}
\end{document}